\documentclass[12pt,preprint]{aastex}
\usepackage{graphicx}
\usepackage{lscape}
\usepackage{color}
\usepackage{txfonts}
\usepackage{natbib}
\usepackage[normalem]{ulem}
\newcommand{\hst}{{\it HST\/}}

\slugcomment{draft version \today}

\shorttitle{HST Large Treasury GO-13297}
\shortauthors{Piotto et al.}

\begin{document}

\def\subr #1{_{{\rm #1}}}

\title{The {\it Hubble Space Telescope} 
UV
Legacy Survey of Galactic Globular Clusters. I. Overview of the Project 
and Detection of Multiple Stellar Populations.
\footnote{Based on
   observations with the NASA/ESA {\it Hubble Space Telescope}, obtained
   at the Space Telescope Science Institute, which is operated by AURA,
   Inc., under NASA contract NAS 5-26555.}}

\author{
G.\ Piotto\altaffilmark{1,2},
A.\ P.\ Milone\altaffilmark{3},  
L.\ R.\ Bedin\altaffilmark{2},
J.\ Anderson\altaffilmark{4},
I.\ R.\ King\altaffilmark{5}, 
A.\ F.\ Marino\altaffilmark{3}, 
D.\ Nardiello\altaffilmark{1,2},  
A.\ Aparicio\altaffilmark{6, 7},
B.\ Barbuy\altaffilmark{8},
A.\ Bellini\altaffilmark{4},
T.\ M.\ Brown\altaffilmark{4},
S.\ Cassisi\altaffilmark{9},
A.\ M.\ Cool\altaffilmark{10},
A.\ Cunial\altaffilmark{1,2},
E.\ Dalessandro\altaffilmark{11},       
F.\ D'Antona\altaffilmark{12},         
F.\ R.\ Ferraro\altaffilmark{11},           
S.\ Hidalgo\altaffilmark{6,7},           
B.\ Lanzoni\altaffilmark{11},                   
M.\ Monelli\altaffilmark{6,7},                  
S.\ Ortolani\altaffilmark{1,2},          
A.\ Renzini\altaffilmark{2},           
M.\ Salaris\altaffilmark{13},          
A.\ Sarajedini\altaffilmark{14},       
R.\ P.\ van der Marel\altaffilmark{4}, 
E.\ Vesperini\altaffilmark{15},        
M.\ Zoccali\altaffilmark{16,17}.          
}

\altaffiltext{1}{Dipartimento di Fisica e Astronomia ``Galileo
  Galilei'', Universit\`a di Padova, Vicolo dell'Osservatorio 3,
  I-35122 Padova, Italy; [giampaolo.piotto;sergio.ortolani]@unipd.it;
  [andrea.cunial;domenico.nardiello]@studenti.unipd.it}

\altaffiltext{2}{INAF-Osservatorio Astronomico di Padova, Vicolo
  dell'Osservatorio 5, I-35122 Padova, Italy;
  [luigi.bedin;alvio.renzini]@oapd.inaf.it}

\altaffiltext{3}{Research School of Astronomy and Astrophysics, The
  Australian National University, Cotter Road, Weston, ACT, 2611,
  Australia; [milone;amarino]@mso.anu.edu.au}

\altaffiltext{4}{Space Telescope Science Institute, 3700 San Martin
  Drive, Baltimore, MD 21218, USA;
  [jayander;bellini;tbrown;marel]@stsci.edu}

\altaffiltext{5}{Department of Astronomy, University of Washington,
  Box 351580, Seattle, WA 98195-1580; king@astro.washington.edu}

\altaffiltext{6}{Instituto de Astrof\`\i sica de Canarias, E-38200 La
  Laguna, Tenerife, Canary Islands, Spain;
  [antapaj;shidalgo;monelli]@iac.es}

\altaffiltext{7} {Department of Astrophysics, University of La Laguna,
  E-38200 La Laguna, Tenerife, Canary Islands, Spain;
  [antapaj;shidalgo;monelli]@iac.es}

\altaffiltext{8}{Universidade de S˜ao Paulo, IAG, Rua do Mat˜ao 1226,
  Cidade Universit´aria, S˜ao Paulo 05508-900, Brazil;
  barbuy@astro.iag.usp.br}

\altaffiltext{9}{Osservatorio Astronomico di Teramo, Via Mentore
  Maggini s.n.c., I-64100 Teramo, Italy; cassisi@oa-teramo.inaf.it}

\altaffiltext{10}{Department of Physics and Astronomy, San Francisco
  State University, 1600 Holloway Avenue, San Francisco, CA 94132,
  USA; cool@sfsu.edu}

\altaffiltext{11}{Dipartimento di Fisica e Astronomia, Universit\`a
  degli Studi di Bologna, Viale Berti Pichat 6/2, I-40127 Bologna,
  Italy;
  [emanuele.dalessandr2;francesco.ferraro3;barbara.lanzoni3]@unibo.it}

\altaffiltext{12}{INAF-Osservatorio Astronomico di Roma, Via Frascati
  33, I-00040 Monteporzio Catone, Roma, Italy;
  dantona@oa-roma.inaf.it}

\altaffiltext{13}{Astrophysics Research Institute, Liverpool John
  Moores University, Liverpool Science Park, IC2 Building, 146
  Brownlow Hill, Liverpool L3 5RF, UK; ms@astro.livjm.ac.uk}

\altaffiltext{14}{Department of Astronomy, University of Florida, 211
  Bryant Space Science Center, Gainesville, FL 32611, USA;
  ata@astro.ufl.edu}

\altaffiltext{15}{Department of Astronomy, Indiana University,
  Bloomington, IN 47405, USA; evesperi@indiana.edu}

\altaffiltext{16}{Instituto de Astrof\'\i sica, Pontificia Universidad
  Cat\'olica de Chile, Av. Vicu\~na Mackenna 4860, Macul, Santiago de
  Chile mzoccali@astro.puc.cl}

\altaffiltext{17}{Millennium Institute of Astrophysics, Av. Vicu\~na
  Mackenna 4860, Macul, Santiago de Chile mzoccali@astro.puc.cl}

\begin{abstract}
  In this paper we describe a new UV-initiative \hst\/ project
  (GO-13297) that will complement the existing F606W and F814W
  database of the ACS Globular Cluster (GC) Treasury by imaging most
  of its clusters through UV/blue WFC3/UVIS filters F275W, F336W and
  F438W.  This "magic trio" of filters has shown an uncanny ability to
  disentangle and characterize multiple-population (MP) patterns in
  GCs in a way that is exquisitely sensitive to C, N, and O abundance
  variations.  Combination of these passbands with those in the
  optical also gives the best leverage for measuring helium
  enrichment.  The dozen clusters that had previously been observed in
  these bands exhibit a bewildering variety of MP patterns, and the
  new survey will map the full variance of the phenomenon.  The
  ubiquity of multiple stellar generations in GCs has made the
  formation of these cornerstone objects more intriguing than ever; GC
  formation and the origin of their MPs have now become one and the
  same problem.  In the present paper we will describe the data base
  and our data reduction strategy, as well as the uses we intend to
  make of the final photometry, astrometry, and proper motions.  We
  will also present preliminary color-magnitude diagrams from the data
  so far collected.  These diagrams also draw on data from GO-12605
  and GO-12311, which served as a pilot project for the present
  GO-13297.
\end{abstract}


\section{Introduction}
\label{intro}

It was once thought that we understood how globular clusters (GCs)
formed in the early Universe, but recent discoveries of their multiple
populations (MPs) have unraveled our long-held view of GCs as ``simple
stellar populations'' formed in a single burst.  It now appears that
virtually {\it all} of them host multiple stellar generations, with
secondary populations characterized by puzzling chemical compositions.
After almost a decade since the first breakthrough (Bedin et al.\
2004), our current understanding can be summarized as follows:

\begin{itemize}

\item {Ubiquity. }  Almost all GCs that have been adequately studied
  are composed of distinct stellar populations (Piotto et al.\ 2012),
  from two in number up to six or more ($\omega$ Cen, Bellini et al.\
  2010).  Even clusters not yet sufficiently probed photometrically
  exhibit the telling Na-O anti-correlation in the spectra of their
  giant stars (Carretta et al.\ 2009).  An exception is
  Rup\,106. Villanova et al.\ (2013) have analyzed spectra of nine RGB
  stars, and concluded that this GC hosts a single stellar population.
  In addition, Walker et al.\ (2011) claimed that IC\,4499 does not
  show evidence of multiple populations. In both cases, higher
  precision stellar photometry is needed for a definitive conclusion
  on the nature of the stellar populations in these two GCs.

\item {Enrichment.}  While the first-generation (1G) stars have a
  composition that can be ascribed to the proto-galactic interstellar
  matter out of which they formed, the second-generation (2G) stars
  can be depleted in C and O, enhanced in N and Na (e.g., Carretta et
  al.\ 2009, Marino et al.\ 2008, 2011a 2011b, 2012 2014), and
  enhanced in He (Milone et al.\ 2012a, 2012b, 2014a).  In some cases
  the He enhancement can be quite strong, close to Y=0.40 (e.g.,
  D'Antona and Caloi 2004, Norris 2004, Piotto et al.\ 2005, 2007,
  King et al.\ 2012, Bellini et al.\ 2013a), while in others it is
  still measurable but small (e.g., $\sim$0.01 as in the cases of NGC
  6397, Milone et al.\ 2012b, and NGC 288, Piotto et al.\ 2013). This
  chemical pattern indicates that the material out of which GCs formed
  was exposed to proton-capture processes at high temperatures and
  must have come from relatively massive 1G stars.  With a few notable
  exceptions, i.e., Omega Centauri (Norris \& Da Costa 1995), M22
  (Marino et al.\ 2009, 2011c), M54 (Carretta et al.\ 2010), NGC 1851
  (Yong and Grundahl 2008), M2 (Yong et al.\ 2014), NGC5824 (Da Costa
  et al.\ 2014), and Terzan 5 (Ferraro et al.\ 2009a), 2G stars have
  the same abundances of iron and other heavy-elements as the 1G
  stars, indicating that their material was somehow not tainted by 1G
  supernova products.  Even in the case of clusters exhibiting
  multiple iron abundances, the differences are relatively small,
  indicating that only a tiny fraction of the supernova products from
  1G were incorporated in 2G stars (Renzini 2013).  It is not clear at
  present whether the clusters with iron abundance variations are part
  of a different class of GCs.  Because of the large number of GCs
  sampled, GO-13297 will surely help to address this idea.

\item {Variety.}  Thus far, we have found no two clusters that
  manifest the MP phenomenon in the same way.  The relative proportion
  of 1G to 2G stars differs enormously from one GC to another, and in
  several cases the 2G stars outnumber 1G stars and are more centrally
  concentrated (as in $\omega$ Centauri, Bellini et al.\ 2010, and 47
  Tuc, Milone et al.\ 2012a, see also Lardo et al.\ 2011, but based on
  lower quality SDSS data.), becoming quite dominant at the very
  center.  In other cases, large differences are also seen in the
  degree of He enhancement and in p-capture elements.
	
\item{Discreteness.}  One crucial property of multiple populations is
  that in a large number of GCs the populations can be separated into
  quite distinct sequences within each CMD and/or in appropriate
  two-color plots, as opposed to a continuous spread, observational
  results that any formation scenario must be able to account for.

\end{itemize}

Different progenitors of the 2G have been proposed. Intermediate-mass
asymptotic giant branch (AGB-M) stars (D’Antona et al.\ 2002),
fast-rotating massive stars (FRMS) (Decressin et al.\ 2007), or
interacting massive binary stars (de Mink et al.\ 2009, Vanbeveren et
al.\ 2012) may eject materials with composition similar to that of 2G
stars, i.e., somewhat helium enriched and exposed to proton capture
reactions at high temperatures, but with sizable differences from one
case to another.\\ The AGB-M and the FRMS models are those for which
all the different aspects of the study of multiple populations have
been more extensively investigated.  Although these models are based
on different sources of processed gas, in both models the amount of
gas available for 2G formation is only a small fraction of the initial
1G.  This has the important implication that both these models require
the 1G star cluster to be initially more massive than it is now (see
e.g. Decressin et al.\ 2007, Renzini 2008, 2013, D'Ercole et al.\
2008, Ventura et al.\ 2014 for estimates of the possible initial 1G
cluster masses), and must have undergone a significant loss of 1G
stars (see, e.g., D'Ercole et al.\ 2008, Decressin et al.\ 2010)
leading to the large fraction of 2G stars currently observed in many
GCs (Renzini 2008).  This large loss of 1G stars also implies that GCs
must have significantly contributed to the formation of the Galactic
halo (see e.g. Vesperini et al.\ 2010, Schaerer \& Charbonnel 2011).
Another requirement shared by these models is that, not only ejected
material from the 1G stars is necessary.  It may have been somehow
diluted with gas with the same composition as the 1G in order to
reproduce the Na-O anticorrelation (as advocated by Decressin et al.\
2007, D'Ercole et al.\ 2008, 2011), though the modalities of such
dilution remain obscure.  The origin of this pristine gas is still a
matter of investigation and debate.  Curiously, recent surveys of
young (10 Myr - 1 Gyr) extragalactic massive clusters (Bastian et al.\
2013a) do not show any evidence of presence of gas and/or ongoing star
formation, though the facts are not quite clear on this (Vink et al.\
2009).  Recently Bastian et al.\ (2013b), following an earlier
suggestion by D’Antona et al.\ (1983), proposed a model which would
not require a larger 1G cluster mass and is based on the accretion of
enriched material released from interacting massive binaries and
rapidly rotating stars onto their circumstellar disks, and ultimately
onto the young pre-main sequence stars.  This model requires some very
ad hoc assumptions (see Cassisi \& Salaris 2014, Salaris \& Cassisi
2014, D'Antona et al.\ 2014) both regarding the lifetimes of the
pre-main sequence accretion disk and regarding the He content of the
accreting material.  Most importantly, the accretion scenario has
serious difficulties to account for the discreteness of the sequences
we observe in the CMDs.

\noindent
In summary, the series of events that led from massive gas clouds in
the early Universe to the GCs we see today remains an intriguing
puzzle.  Clearly, GC formation was a far more complex process than
ever imagined before.  Clues and insight needed to solve this puzzle
can come only from homogeneous mapping of a large sample of GCs with
appropriate multi-band photometry.  A photometric survey can map a
large sample of stars, at all evolutionary phases.  Such a survey will
enable us to identify 2G stars even in the cases where they represent
a minor fraction of the entire cluster population.  Multi-band
photometry allows us to infer chemical properties of stars in a
complementary way to spectroscopy (Milone et al.\ 2012a), and to do
this for a large sample of GCs.  This is the subject of the GO-13297
HST Large Treasury project (PI G.\ Piotto) approved for Cycle 21.
This paper will describe in detail the project, the selected sample of
clusters (see Table 1), the observational strategy, the data-reduction
strategy, and the intended use of the data by the team.  We will also
present preliminary results from the data acquired thus far.

\section{The magic of WFC3's UV filters.} 
\label{magicUVs}

\begin{figure}[!ht]
\epsscale{0.8}
\plotone{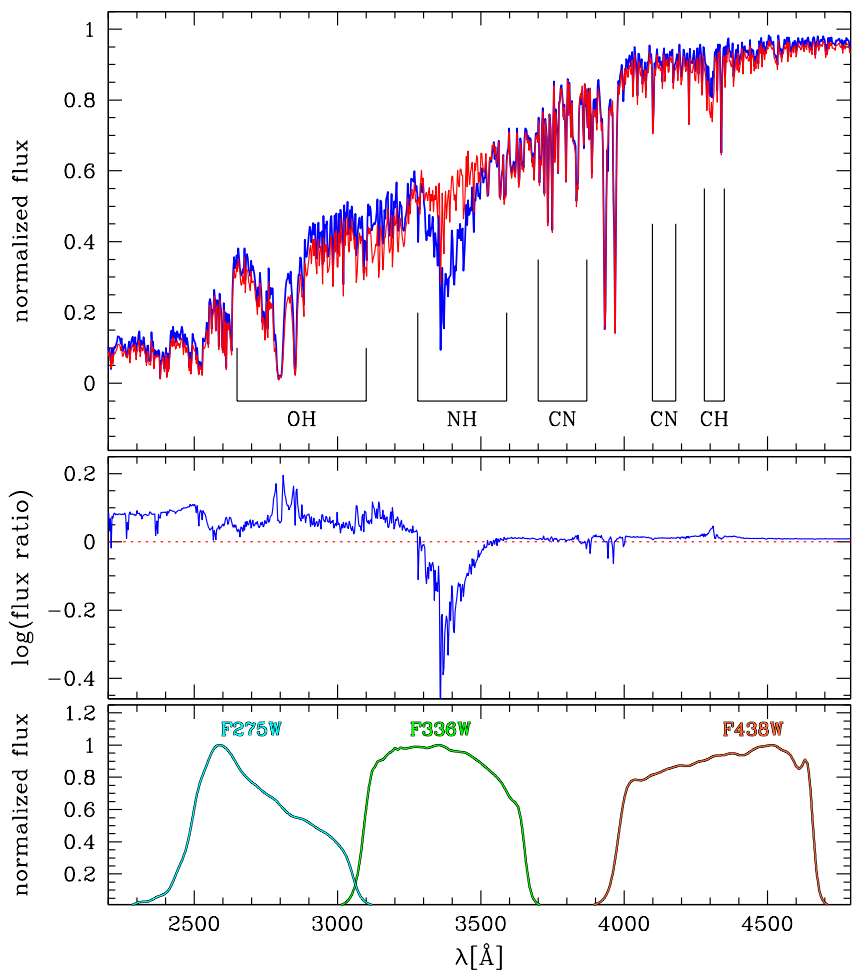}
\caption{{\it Upper panel:} In red the simulated spectrum of a star
of the first stellar generation (N-poor) RGBa in NGC~6752; in blue the
simulated spectrum of a third generation, N-rich, RGBc star (Milone et
al.\ 2010). {\it Middle panel:} Flux ratio of the two spectra
reproduced in the upper panel. {\it Lower panel:} Bandpasses of the
F275W, F336W, and F438W WFC3/UVIS camera.}
\label{magicUVf}
\end{figure}

Previous studies, based on spectroscopy and ground-based photometry
have shown that Str\"oemgren and ultraviolet photometry does a good
job in separating multiple populations along the RGB on nearbly GCs
(e.g. Yong et al.\,2008; Marino et al.\,2008).  Our WFC3/UVIS
observations in Cycles 18--20 have opened a spectacular new window by
showing that UV/optical colors separate the various sub-populations
much better than the traditional optical colors alone (Bellini et al.\
2010, 2013a, Milone et al.\ 2010, 2012a, 2012b, Piotto et al.\ 2012,
2013) and do so across the {\it entire\,} color-magnitude diagram,
from the main sequence (MS) all the way to the horizontal branch (HB).
Figure~\ref{magicUVf} shows the power of the UV photometry for the
case of NGC 6752.  It reproduces the simulated spectra of RGBa
(N-poor, He-poor red line, first stellar generation ) and RGBc
(N-rich, He-rich blue line, third stellar generation) of NGC~6752
(Milone et al.\ 2010).

The reason why F275W, F336W, and F438W work so well is quite simple.
Milone et al.\ (2012a) emphasized that the F275W passband includes an
OH molecular band, F336W an NH band, and F438W (or F435W for ACS) CN
and CH bands, as illustrated in Fig.~1. This property of the HST
filter system is at the basis of the project we are presenting in this
paper.  The 1G stars, which are oxygen- and carbon-rich and
nitrogen-poor, are relatively faint in F275W and F438W, but bright in
F336W.  Conversely, 2G stars, whose material has been CNO-cycle
processed, are oxygen- and carbon-poor but nitrogen-rich. As a
consequence, they are relatively bright in F275W and F438W but faint
in F336W.  Therefore, 1G stars are bluer than 2G stars in one color
(F336W $-$ F438W), but redder in another (F275W $-$ F336W), and this
{\it inversion} is seen in Fig.~\ref{magicUVf} Milone et al.\ (2013)
defined a pseudo-color $C_{\rm F275W,F336W,F438W}$ =($m_{\rm
  F275W}-m_{\rm F336W}$)$-$($m_{\rm F336W}-m_{\rm F438W}$), which
maximizes the virtue of both F336W $-$ F438W and F275W $-$ F336W, and
has proven to be quite efficient in separation of multiple sequences.

\begin{figure}[!ht]
\epsscale{0.8}
\plotone{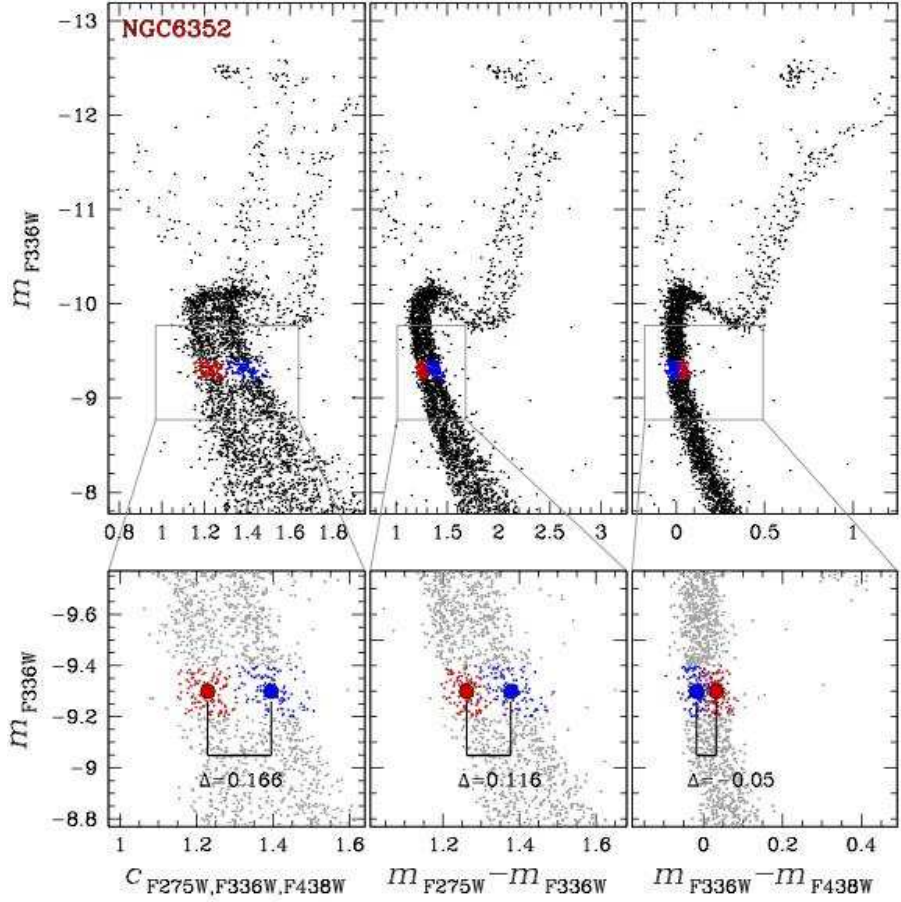}
\caption{CMDs of NGC 6352 from GO-13297 data.  $m_{\rm F336W}$ vs.\
  $C_{\rm F275W,F336W,F438W}$ (left panels), $m_{\rm F336W}$ vs.\
  $m_{\rm F275W}-m_{\rm F336W}$ (middle panels), and $m_{\rm F336W}$
  vs.\ $m_{\rm F336W}-m_{\rm F438W}$ (right panels). Lower panels show
  a zoom of the MS.  Red dots highlight 2G stars; blue dots 1G stars.}
\label{n6352}
\end{figure}

Figure~\ref{n6352} shows an example of the application of GO-13297
data to NGC 6352.  Left panels show the power of the pseudo-color
$C_{\rm F275W,F336W,F438W}$ in separating the two MSs.  The middle
panels show that in the $m_{\rm F336W}$ vs.\ $m_{\rm F275W}-m_{\rm
  F336W}$ color-magnitude diagrams (CMDs), 1G stars are redder than 2G
ones, but they become bluer in the $m_{\rm F336W}$ vs.  $m_{\rm
  F336W}-m_{\rm F438W}$ CMD (right panels).

Thanks to F275W, F336W, and F438W data, the two stellar generations
are clearly distinguished and are seen to wind, in a distinct but
intertwined fashion, from the MS through the subgiant branch (SGB) and
red-giant branch (RGB), to the HB.

The UV passbands are thus able to probe the CNO content of the cluster
stars.  Note that the nuclear processes that produce the light-element
(anti-)correlations (e.g. the Na-O or the Mg-Al anticorrelations) also
produce helium, and measuring the He enhancement of each
sub-population is particularly important to help pinpoint what kind of
stars may have produced the 2G material.  The photometric impact of He
is mainly on the optical bands (through the stellar temperature).
After population-tagging stars according to their UV colors, we can
make fine distinctions ($\sim$0.01 mag in color) in the MS ridge lines
of the generations in the optical bands, and thus infer variations of
only a few per cent in He (Milone et al.\ 2010, 2012ab).  In summary,
the combination of UV and optical colors allows us to estimate the
content of C, N, O, and He, even for stars that are far too faint to
pursue with spectroscopy.  (See Milone et al.\ 2012ab, 2013 for
applications of the method.)

The combination of UV/blue filters that we have selected for GO-13297
is also ideally suited for studies of HB stars, blue stragglers stars
(BSSs), and compact binaries.  For hot HB stars, optical colors
quickly become degenerate, but UV colors are able to follow the entire
HB extension.  This is particularly interesting for those GCs (even
some metal-rich ones, Rich et al.\ 1997, Busso et al.\ 2007, Bellini
et al.\ 2013a) that have very extended HBs and often exhibit the most
complex MPs.  Enhanced helium also distorts the HB morphology:\ when
$Y$ is increased from 0.23 to 0.38 the blue HB becomes 1 mag brighter
in F275W, while the EHB becomes 0.5 mag fainter.  Thus the UV band is
crucial for determining the physical parameters of HB stars (including
$T_{\rm eff}$) and their helium content (Rood \& Crocker 1989,
D'Antona and Caloi 2008).  BSSs, compact binaries, and high-energy
sources are also best studied with filters of shorter wavelength.

\section{The need for a broad survey}
\label{broadsurvey} 
Detailed studies, made possible largely by \textit{HST}, have revealed
remarkable variety and complexity.  As mentioned above, among the
dozen or so GCs in which multiple photometric sequences have been
resolved thus far, the fraction of 2G stars ranges from a few per cent
to well over 50\%.  Each GC has its own individuality, thus pointing
to a great deal of variance in the formation process itself.  Yet the
sheer range of variance remains to be established, and cannot be
properly mapped without the widest possible survey of Galactic GCs.
Furthermore, since spectroscopy has shown that nearly all GCs host
MPs, it is all the more appropriate to move from studying a few ``odd
ducks'' to a systematic survey of the whole flock.
 
The largest homogeneous set of \textit{HST}\, data on GCs until now is
the ACS Treasury database of F606W and F814W images for 65 GCs
(GO-10775, PI Sarajedini, Anderson et al.\ 2008); however, its
optical-wavelength baseline offers sub-optimal leverage for MP
studies.  The new GO-13297 data set represents what is needed:\ to
augment the existing database with UV and blue observations through
the ``magic trio'' of WFC3/UVIS filters, F275W, F336W, and F438W,
whose power in population separation we have just illustrated.  Our
photometry will characterize the populations in each cluster:\ how
many stellar generations are present, and in what relative proportions
--- even for small minority populations that might easily escape
spectroscopic surveys.  The UV/blue WFC3/UVIS photometry, combined
with visible photometry from the ACS GC Treasury, will also enable a
systematic survey and detailed intercomparisons of the populations of
EHB stars, BSSs, and compact binaries in a large sample of Galactic
GCs, as outlined in Section~\ref{legacy}.

\section{The GC Sample and the Observing Strategy}
\label{sample}

Figures 1 and 2 show that multiple populations of many clusters can be
readily disentangled by means of a two-color diagram made from F275W,
F336W, and F438W.  In order to effect a clean separation of the
populations with the CMDs and the pseudo-color diagrams, our
observational goal was a color precision of about 0.02 magnitude or
better, which implies a $S/N$ of 50 or better in F275W (the most
photon-starved filter).  We did not try to follow the multiple
sequences all the way down the main sequence for every cluster ---
that would have been prohibitively expensive orbit-wise.  Rather, our
goal was simply to maintain this precision in F275W to just below the
turn-off.  That will naturally give us a population separation for the
evolved populations as well.

Given that we need a minimum of 4 exposures in each band to make a
robust catalog, two orbits represent the minimum set of observations
for each cluster.  Fifteen clusters are observed with that allocation
(see Table~1 for a detailed description of the observing time
allocated for each cluster and for the different filters).  Four
clusters need 3 orbits each, eight clusters need 4 orbits, three need
5 orbits, and two clusters need 6 orbits each.  (One of the 6-orbit
clusters is NGC 6715 [M54], which is too far away to get
0.02-magnitude precision below the turnoff in F275W.  But it is likely
to have a larger MS spread, and therefore not need a $S/N$ as high as
50.  Furthermore, understanding its upper sequences with a
MP-sensitive filter system will be important.)

The remaining 14 clusters are so far away or too reddened that we
cannot hope to resolve MS splits with F275W; for these clusters we
allot two orbits, which are adequate for the SGB, RGB, and HB
populations.

We observe each GC in two visits, with orientations separated by
roughly 90 degrees, to allow a direct validation of the
charge-transfer efficiency (CTE) corrections (see below).  For most
GCs, we take two exposures per filter per visit.  These exposures are
dithered to cover the gap, so that most stars will be found in four
exposures in each band, thus allowing a robust empirical estimate of
the photometric errors.

In summary (see Table~1), the GO-13297 GC sample includes 47 GCs and a
very old ($\sim8$ Gyr, King et al.\ 2005), open cluster, NGC 6791,
which has been included in the sample because of its many
peculiarities (Bedin et al.\ 2008, Geisler et al.\ 2012).  The program
has been allocated 131 orbits.  Originally, the program also included
NGC 4147. Unfortunately, the cluster could not be scheduled (for
guiding star problems) and has been substituted with M4.  Before
GO-13297 we had a pilot project, GO-12605 (PI Piotto, 22 orbits),
where we began our collection of WFC3/UVIS F275W, F336W, and F438W
images for seven GCs (M3, M13, M15, M80, NGC 288, NGC 362, and NGC
2808), extending previous F606W and F814W data from GO-10775.  In an
even earlier project, GO-12311 (PI Piotto, 17 orbits), we had
collected WFC3/UVIS F275W images in the GO-10775 fields of five more
clusters (47Tuc, M4, M22, NGC 1851, and NGC 6752).  The GO-12311 data
set is now complemented by F336W and F438W WFC3/UVIS images for NGC
1851, M4, and M22.

Some of the GCs from the previous GO-10775 Treasury program were not
included in the GO-13297 target list either because they are too
poorly populated for a meaningful search for multiple sequences, or
because they are highly reddened or extremely distant, which would
have made them impossible to observe in the UV in a reasonable number
of orbits.  In summary, completion of the GO-13297 program gives us,
across all programs, 57 GCs with a homogeneous set of images in the
five photometric bands F275W, F336W, F438W, F606W, and F814W.

\begin{figure}[!ht]
\epsscale{0.8}
\plotone{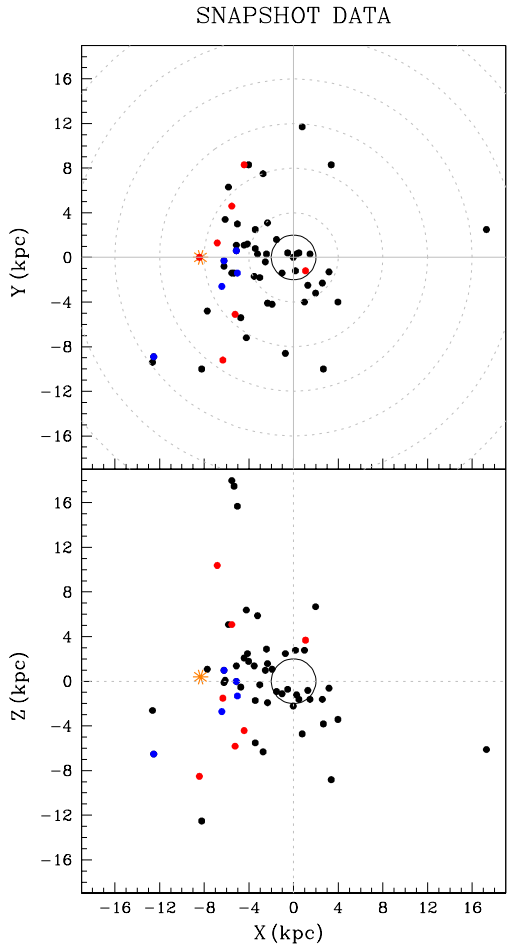}
\caption{Spatial distribution of the target clusters for GO-13297
  (black), GO-12605 (red), and GO-12311 (blue), in Galactic $XYZ$
  coordinates. The adopted Sun position is indicated by an asterisk. }
\label{GCdist}
\end{figure}

Table 1 lists these 57 clusters and summarizes their WFC3/UVIS and
parallel ACS images.  The final five columns in the Table give the
Galactic $XYZ$ coordinates, the reddening, and the distance modulus
(from the Harris 1996 catalog, 2010 edition).  GCs observed within
GO-12311 are marked by an asterisk; GCs from GO-12605 by two
asterisks.  Figure \ref{GCdist} shows the distribution of the clusters
in the Galactic $XY$ and $XZ$ planes.

WFC3/UVIS has been on board \textit{HST}\, for only a few years, yet
CTE losses have become a concern for it much more quickly than they
were for ACS.  The UVIS team has traced this to the fact that CTE
losses are particularly bad when the background is low (see MacKenty
\& Smith 2012).  In order to minimize CTE losses, we determined the
anticipated background for each exposure and supplemented it with
postflash to ensure a minimum of 12 e$^-$.  In addition, to the extent
possible, we observed each cluster at two different orientations 90
degrees apart so that any remaining CTE issues can be dealt with
properly.  We will postflash our images to ensure a background of a
least 12 e$^-$.

The main focus of our program is UV imaging of the central field of
each cluster, and our observing strategy was optimized for that field.
Even so, we could still afford to take $\sim$4 medium-length exposures
of a parallel field with ACS (more for the clusters with more than two
orbits).  Our general strategy has been to take one orbit at one
orientation and another at a different orientation.  This resulted in
placing the ACS/WFC camera on two different fields 6\arcmin\ from the
cluster center.  For large He differences among the different
populations, even these few parallel observations will allow us to
detect split MSs, as was successfully done in the case of NGC 2808
(Piotto et al.\ 2007).  The mapping of the outskirts of the clusters,
albeit at lower photometric accuracy, adds a large baseline to the
spatial aspects of our MP study, and will allow us to measure the
1G/2G radial gradient.  The relative concentration of the 2G stars
provides critical information about the formation process (see Bellini
et al.\ 2009 and D'Ercole et al.\ 2008).

When possible, we have chosen an orientation that allows the parallel
fields to overlap previous observations, so that we can also measure
proper motions for the outer field.  At a minimum, we can use the
WFPC2 parallels taken during the GO-10755 program, but many clusters
have even better archival data.  Note that the overlaps with previosly
collected GO-10775 images provide us with a $\sim$7-year temporal
baseline for relative proper-motion measurement.  The outer-field
observations should be deep enough to allow us also to measure
absolute proper motions for the clusters, relative to background
galaxies.

\section{Data Reduction}

The ultimate goal of this program is to provide comprehensive catalogs
to the community in order to enable the many studies that this
57-cluster data set makes possible.  It will not be possible to
produce the best possible catalogs immediately.  Once we have the
entire data set in hand, we will be able to do an exquisite
characterization of the PSFs and distortion solutions for each filter,
as well as empirical corrections for any remaining CTE issues.
Therefore the final catalogs will not be available until about a year
after the last data are taken.

There are, however, many studies that will not require a definitive
catalog, and we are making preliminary products available to the
community as soon as possible.  In addition, team members will be
performing special processing on individual clusters to produce early
science results.

We describe in the following subsections the three different
reductions we envision for this project: the early-release preliminary
reduction, the specialized reductions for early science, and the final
comprehensive reduction.

We have created a web page
(http://groups.dfa.unipd.it/ESPG/treasury.html) where interested
persons can follow the progress of the project and retrieve the
released photometric and astrometric data. A mirror site will also be
available in the MAST archive, at STSCI.

\subsection{Early-release reduction}
\label{dirty}

Much of the progress in understanding the MP phenomenon will be
possible only by combining photometric observations of larger numbers
of stars with targeted spectroscopic studies of individual stars.  In
an effort to allow spectroscopists to identify stars and plan their
runs as soon as possible, we will provide an early reduction of the
data shortly after the publication of this paper. We will update the
archive as more data come in.

The preliminary reduction will focus on measuring the stars in the
simplest way possible and collating them with the existing GO-10775
catalogs in F606W and F814W.  To this end, we run on each exposure a
version of img2xym\_wfc3uv (similar to the one-pass star finding and
measuring for ACS described in ACS/ISR-2006-01 by Anderson \& King,
but adapted to the WFC3/UVIS detector).  This is a one-pass finding
routine that goes through the image pixel by pixel, identifying as a
potential star any pixel that has more than 100 counts within its
inner 3x3 pixels and has no brighter pixels within a radius of 4
pixels.  On each identified source we then use an empirical library
PSF to perform PSF-based 5$\times$5-pixel aperture photometry and
determine an accurate position.  It then uses the WCS information to
transform the measured positions into the GO-10775 catalog frame and
matches up each star list to the stars in the catalog.

Once this has been done for all the available exposures, we collate
the observations of each star in each filter and produce a catalog
that has the following elements: First, we provide a 2013-2014
position that can be compared against the 2006 position from GO-10775.
Since we have not optimized the PSFs and astrometric transformations,
this position has only 0.1-pixel precision, but that should be
sufficient for cluster-field separation.  We next provide the F275W,
F336W, and F438W photometry for all the stars in the GO-10775 catalog
that could be found by our one-pass routine.  We provide the
VEGA-mag-calibrated average magnitude, the RMS deviation of the
individual observations about this average, and the number of images
that contributed to the average.  Finally, we also produce stacked
images in each of the three new bands, in the frame of the original
F606W products.

\subsection{Data reduction for early-science papers}
\label{presentred}

The early-release catalog described above has been constructed with a
library PSF and a one-pass finding approach that treats each star as
if it is isolated.  The final reduction will improve on several of
these limitations, but since it will be more than a year before the
final reductions are available, in the meantime several members of the
team have been running supplemental reductions in order to start the
detailed scientific analysis of several clusters.

These reductions will be described in more detail in the individual
papers (for example, see Milone et al.\ for M2, submitted to ApJ),
but many of them will make use of photometry software (such as that
described in Bellini et al.\ 2010) that allows for the fact that the
PSFs in individual exposures often differ slightly from the library
PSF.  The specialized reductions will also include
differential-reddening corrections.  Both of these improvements are
necessary to enable the best possible sequence separation for study of
the multiple populations.

Other early-science projects will focus on the internal proper
motions.  The one-size-fits-all preliminary products will not do the
careful transformations and multi-epoch fitting that is necessary to
measure accurate motions and their errors.  For the clusters of
particular interest, we will reduce all the available epochs and
explore the internal motions in papers before the final catalogs are
available.

These data are available for specific projects under request to the
first author of this paper, and will be stored on the dedicated web
page at the Department of Physics and Astronomy, of the University of
Padova (http://groups.dfa.unipd.it/ESPG/treasury.html). The preliminary
photometry coming from this process is at the base of the CMDs we
present in this paper.

\subsection{Final Data Reduction}
\label{magicred}

The final reduction for this program will be possible only after all
of the data are collected.  The full data set will enable us to
construct an exquisite model of both the average PSF and the average
distortion solution in each of the three filters.  We will also
construct a tailored PSF for each exposure to reflect the fact that
breathing causes it to vary slightly from the average.

Whereas for the preliminary reduction (described in
Section~\ref{dirty}, \ref{presentred}) we reduced each exposure
separately with a simple one-pass approach and combined the starlists
with the existing GO-10775 catalog, for the final reduction we will
construct an entirely new catalog using an approach that analyzes all
exposures simultaneously in a multi-pass, artifact-avoiding,
neighbor-subtracting algorithm similar to that used to reduce the
original GO-10775 data set (Anderson et al.\ 2008).  This new approach
will involve all five filters and will produce proper motions,
source-subtracted stacks (which anyone can search for remaining
objects of any sort), and a variety of artificial-star tests.

In addition to reducing the GO-10775 and GO-13297 data in a uniform
way, we will also reduce all of the existing WFPC2, ACS, and WFC3 data
for the clusters and cross-identify the stars with our catalog,
producing time-series photometry and astrometry that can be used for
high-precision proper motions and variable-star studies.  From these
many individual observations of each star, we will measure a
variability index and a high-precision proper motion, along with an
error estimate of thr latter.  Along with each photometric
observation, we will also provide a differential-reddening correction,
so that accurate CMDs can be constructed for multi-population
analysis.

\section{High Level Products}

\subsection{Preliminary Release of Astrometry, Photometry and Proper
  Motions} 

The data products described in Section~\ref{magicred} (the new F275W,
F336W, and F438W photometry for the GO-13297 star catalog and new
stacked images) are being made available via the
http://groups.dfa.unipd.it/ESPG/treasury.html WEB page and the STSCI
MAST archive (via the high-level product link in the archive listing
for the proposal).  We will continually update the files as new
observations come in.

\subsection{The Final Catalogs}

Once all the data are in hand, we will perform the specialized uniform
reductions described in Section~\ref{magicred}.  In addition to
outputting those reductions we will also produce for each cluster
several other high-level products:

\begin{itemize} 

\item A unified star list for the central field for all filters.  We
  will have at a minimum F275W, F336W, F438W, F606W, and F814W for all
  clusters.  We will also link up other relevant photometry from other
  archival data, such as SBC and FUV WFPC2 observations, which are
  particularly relevant to hot populations.  For each filter, we will
  provide an average flux and its error, based on internal agreement.
  We will also provide the flux measured for each star in each
  exposure, so that users can search for variability.

\item A differential-reddening map for each cluster, and a correction
  for each measurement of each star.

\item Proper motions for most of the stars in the catalogs.  These
  motions will be focused on the UV data set, so they will extend from
  the brightest giants to a few magnitudes below the turnoff, so that
  they will be particularly useful for comparison with radial
  velocities and spectra from the ground.

\item Artificial-star tests done for stars inserted along the fiducial
  sequences.  This will allow users to quantify the extent to which
  crowding may broaden the sequences.

\item Stacked images of the field in all filters, co-registered and
  with 2MASS-calibrated WCS headers.  This includes data sets from the
  archive and any other data we can link-up (including Chandra
  images).  We will also provide subtracted stacks, so that the
  community users can determine whether there might be any stars that
  escaped our finding algorithm.

\item A specialized website portal to serve up the data in a way that
  is most useful to globular-cluster researchers.  It will allow users
  to zoom in and out of images and select stars for dynamic display on
  CMDs and PM diagrams.  Stars will be flagged by evolutionary stage
  and 1G/2G designation.  We will also indicate which stars have
  spectra available in the archive that we will maintain.  We will, of
  course, also deliver our catalogs and images to MAST for a more
  standard dissemination.

\end{itemize}

While there is much science we are eager to do on this data set, the
primary focus of the program is the generation of a legacy data set
for the entire community.  That is why we are providing the
preliminary reductions as quickly as possible, and are equally intent
on getting the final high-level products to the community within a
year of the last observation.

\section{First Color-Magnitude Diagrams}

At the time of the submission of the present paper, about 95\% of the
GO-13297 observations have been collected.  Though it will take some
time before we can perform the final reduction as described in
Section~\ref{magicred}, we are constantly monitoring the data
acquisition, and all data are reduced following the procedures
described in Section~\ref{presentred}.  A more detailed analysis of
the multiple stellar population for some specific, particularly
interesting clusters is ongoing, and corresponding papers will be
submitted in the next few months. Here, we want to make public the
first CMDs, as they come out from our preliminary photometry.  The
data reduction process is still ongoing, and the purpose of the CMDs
we are presenting here is just to show the potentiality of the
GO-13297 database.

\begin{figure}[!ht]
\epsscale{1.00}
\plotone{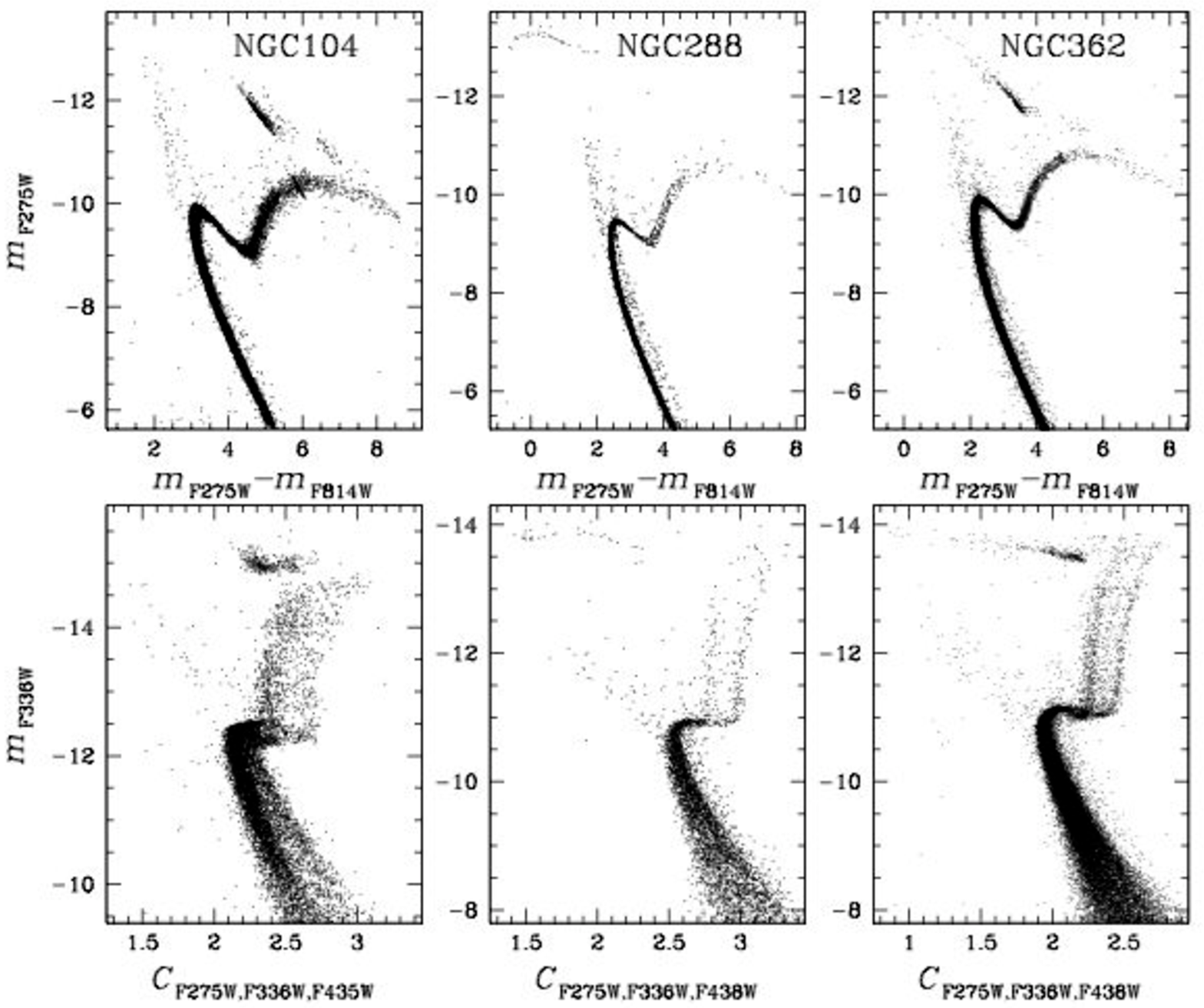}
\caption{\textit{Upper panel} $m_{\rm F275W}$ vs. $m_{\rm F275W} -
  m_{\rm F814W}$ CMD of NGC 104, NGC 288 and NGC 362.  \textit{Lower
    panel}. $m_{\rm F336W}$ vs. $C_{F275W,F336W,F438W}$ index for the
  same clusters as upper panel.  Magnitudes and colors are in the
  instrumental system described in the text.  }
\label{cmd1}
\end{figure}

\begin{figure}[!ht]
\epsscale{1.00}
\plotone{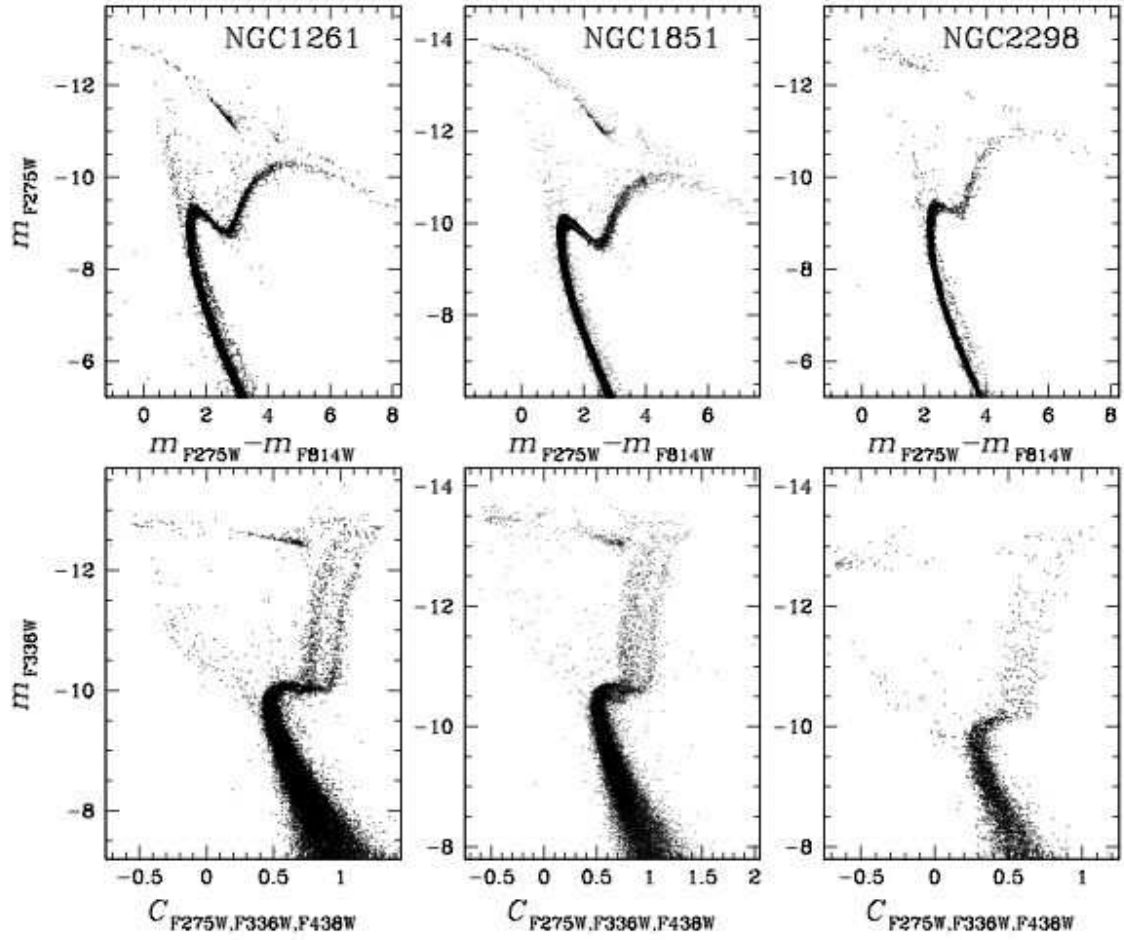}
\caption{As in Fig.~\ref{cmd1}, for NGC 1261, NGC 1851, and NGC 2298.}
\label{cmd2}
\end{figure}

\begin{figure}[!ht]
\epsscale{1.00}
\plotone{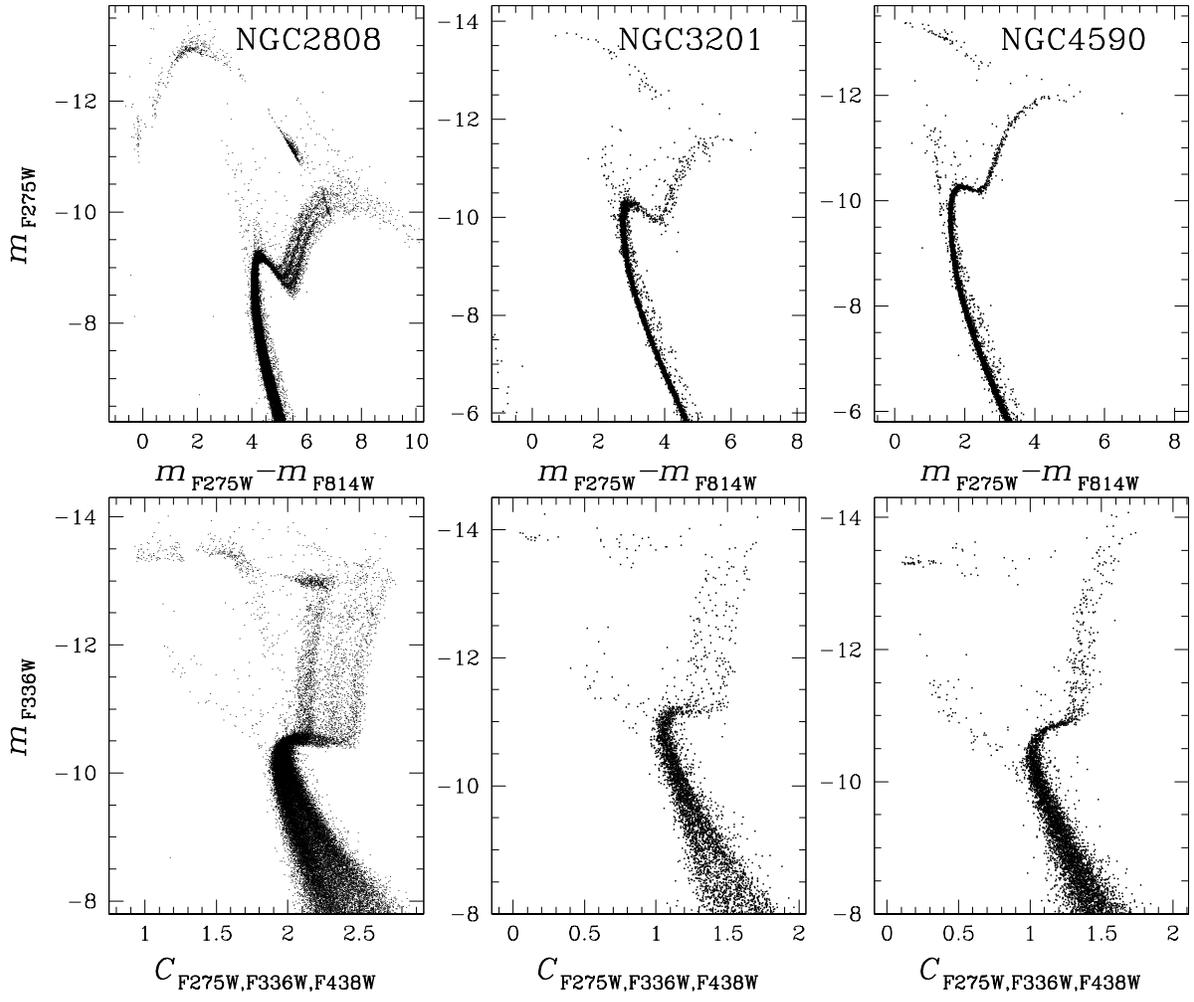}
\caption{As in Fig.~\ref{cmd1}, for NGC 2808, NGC 3201, and NGC 4590.}
\label{cmd3}
\end{figure}

\begin{figure}[!ht]
\epsscale{1.00}
\plotone{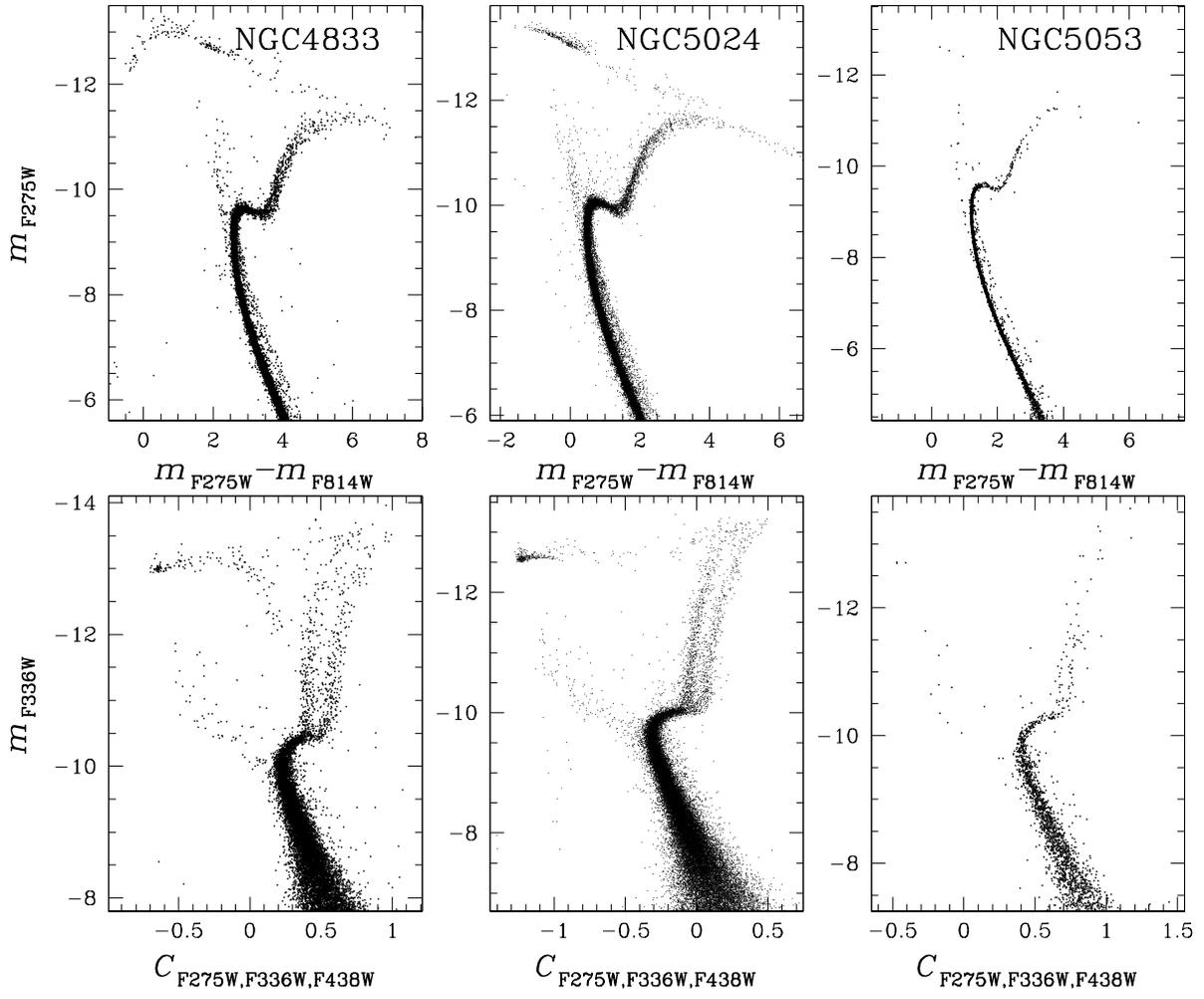}
\caption{As in Fig.~\ref{cmd1}, for NGC 4833, NGC 5024, and NGC 5053.}
\label{cmd4}
\end{figure}

\begin{figure}[!ht]
\epsscale{1.00}
\plotone{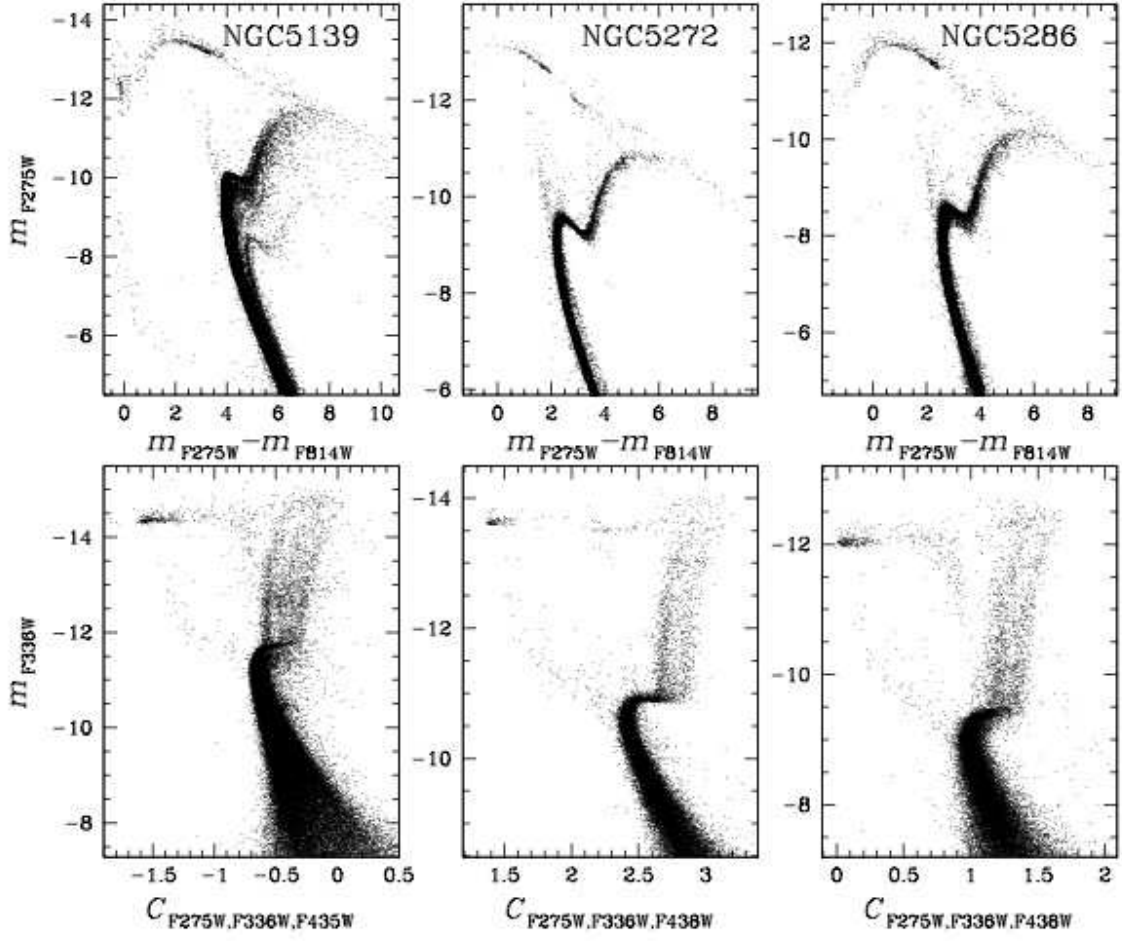}
\caption{As in Fig.~\ref{cmd1}, for NGC 5139, NGC 5272, and NGC 5286.}
\label{cmd5}
\end{figure}

\begin{figure}[!ht]
\epsscale{1.00}
\plotone{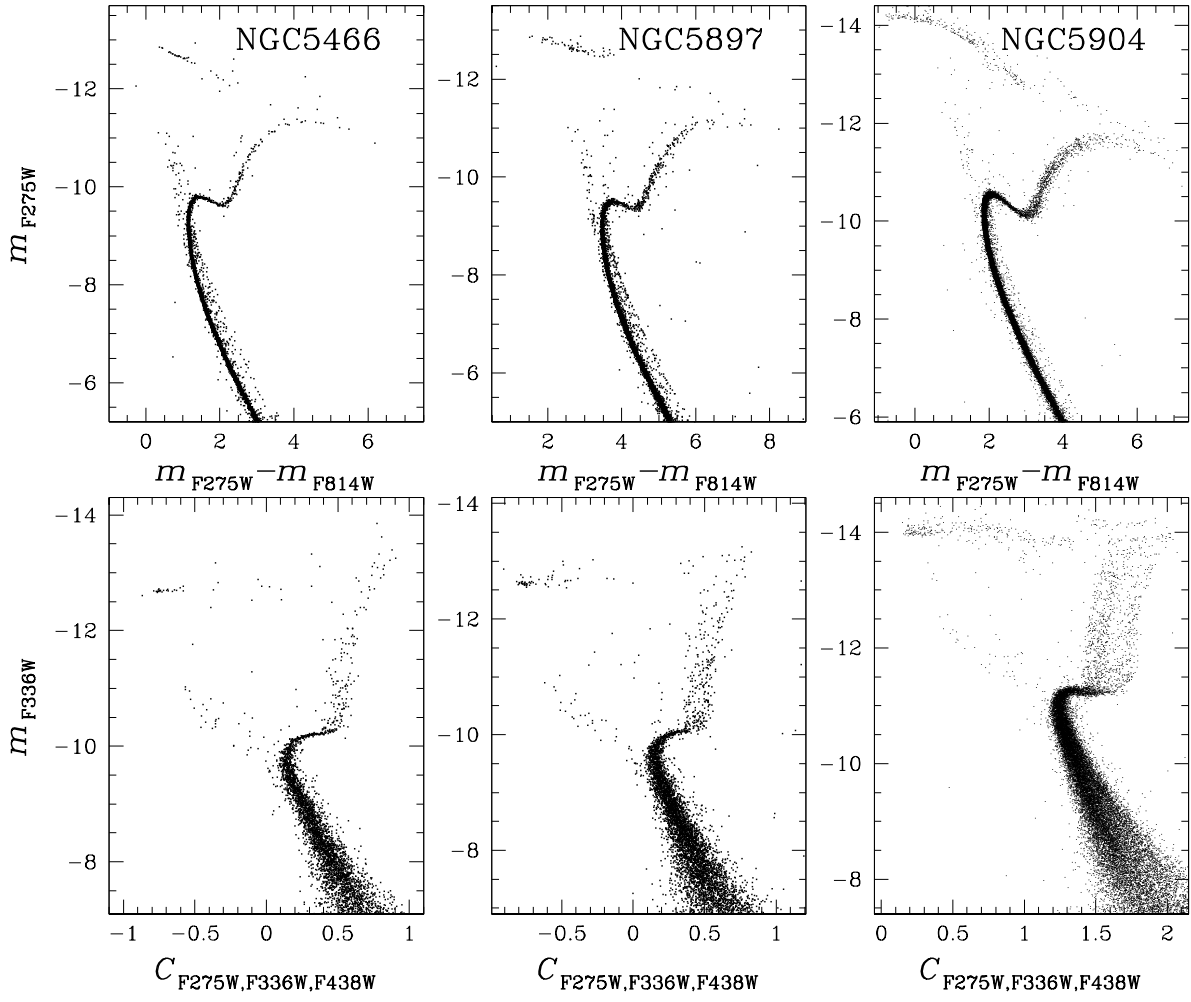}
\caption{As in Fig.~\ref{cmd1}, for NGC 5466, NGC 5897, and NGC 5904.}
\label{cmd6}
\end{figure}

\begin{figure}
\epsscale{1.00}
\plotone{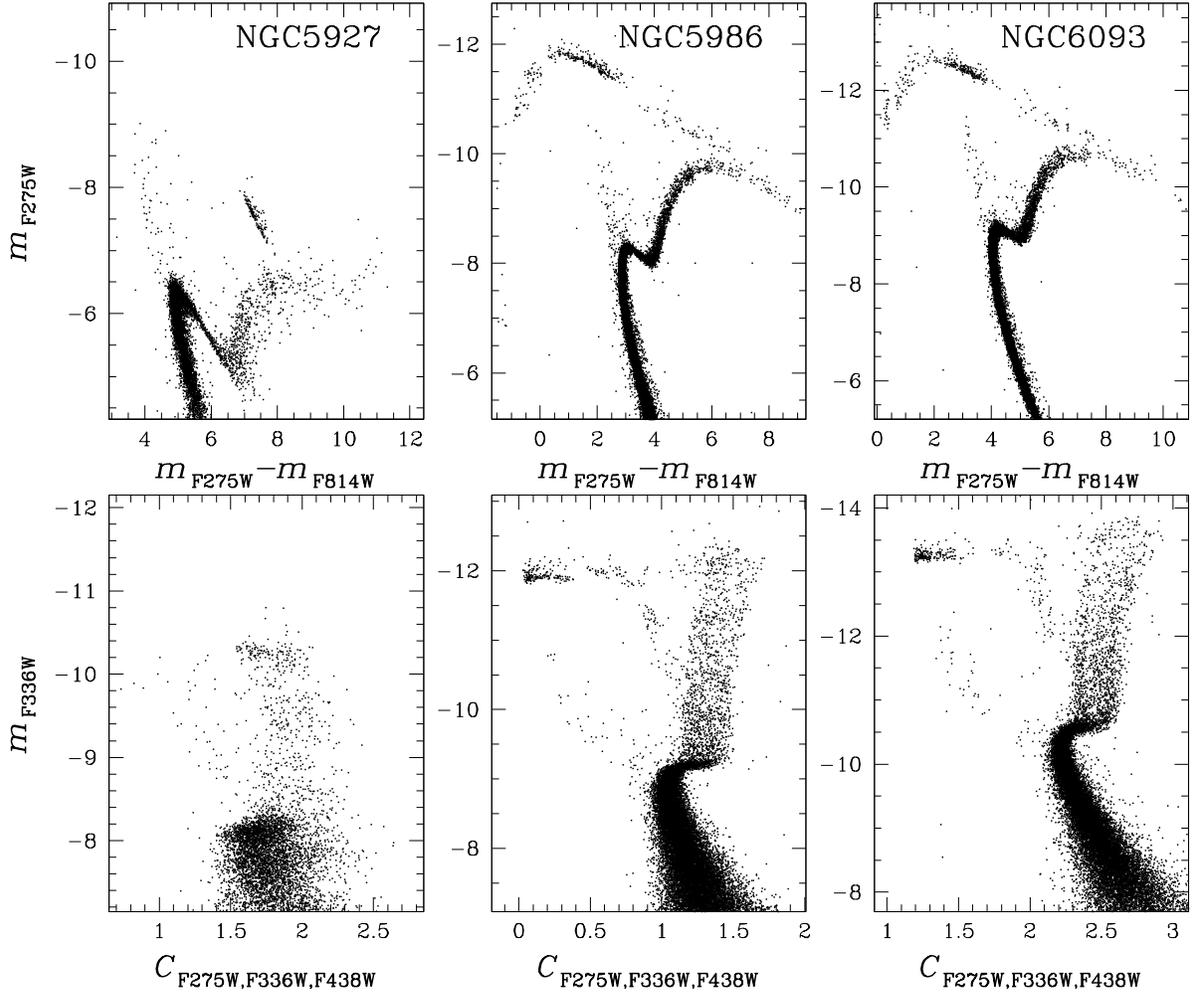}
\caption{As in Fig.~\ref{cmd1}, for NGC 5927, NGC 5986, and NGC 6093.}
\label{cmd7}
\end{figure}

\begin{figure}
\epsscale{1.00}
\plotone{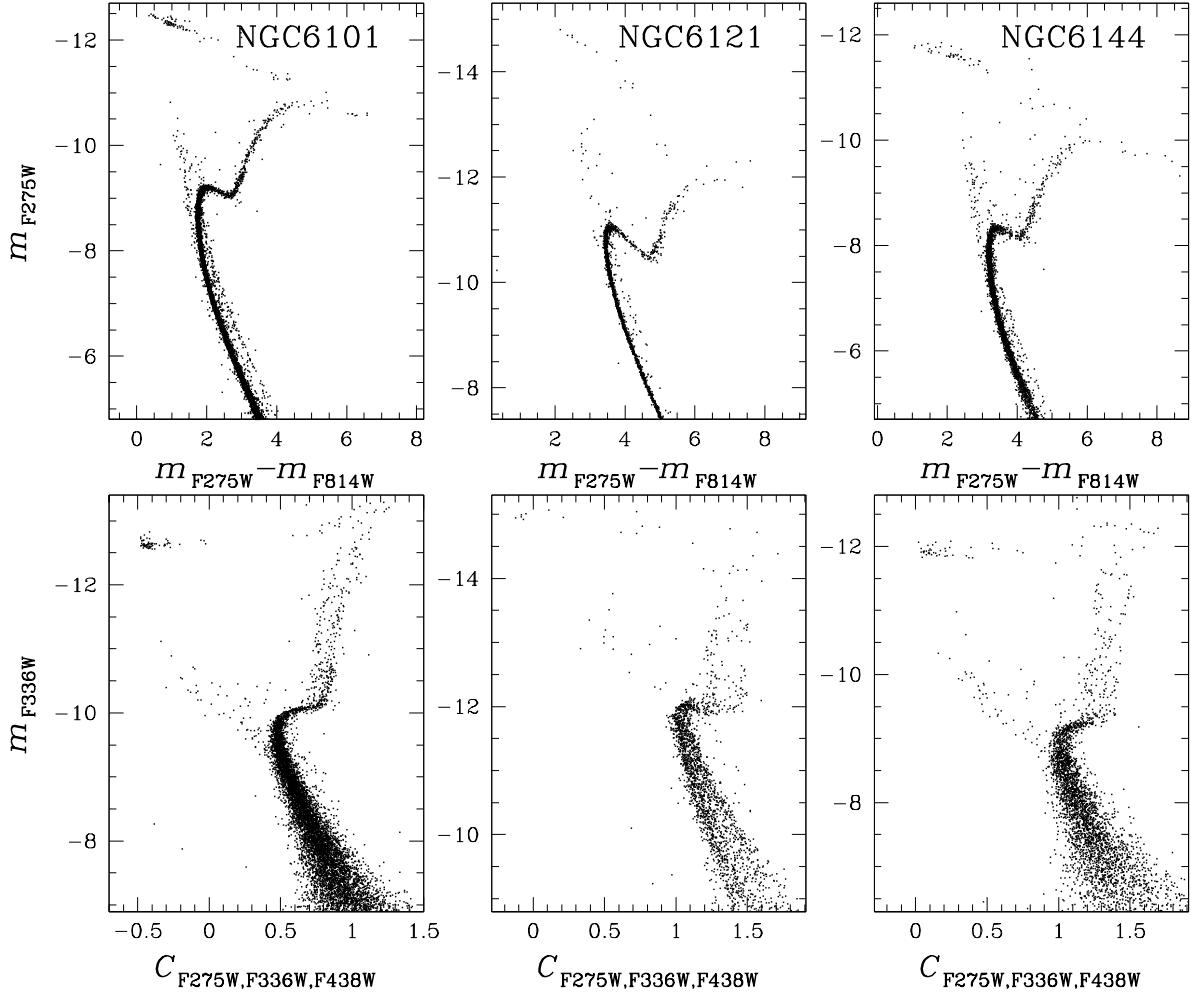}
\caption{As in Fig.~\ref{cmd1}, for NGC 6101, NGC 6121, and NGC 6144.}
\label{cmd8}
\end{figure}

\begin{figure}
\epsscale{1.00}
\plotone{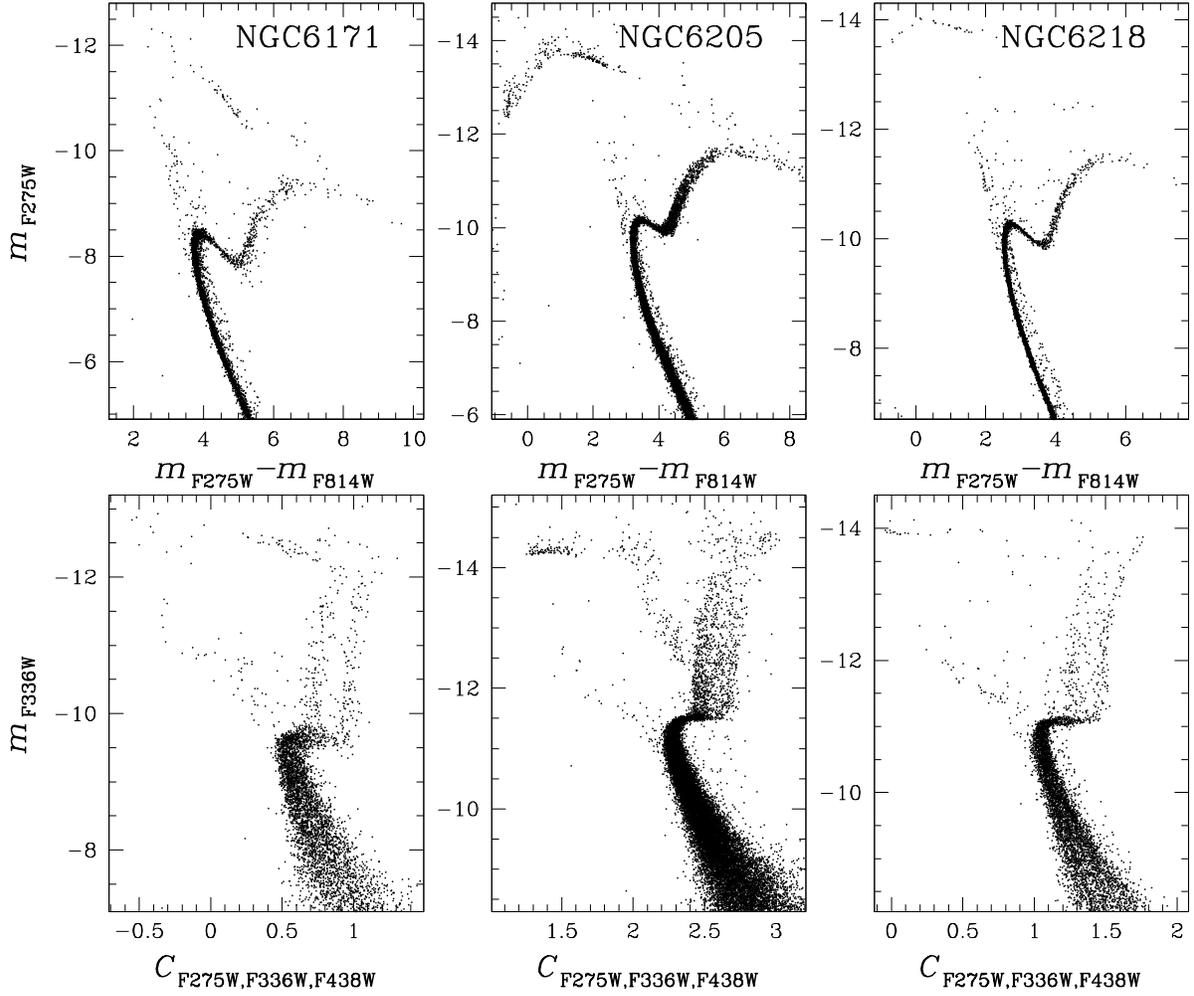}
\caption{As in Fig.~\ref{cmd1}, for NGC 6171, NGC 6205, and NGC 6218.}
\label{cmd9}
\end{figure}

\begin{figure}
\epsscale{1.00}
\plotone{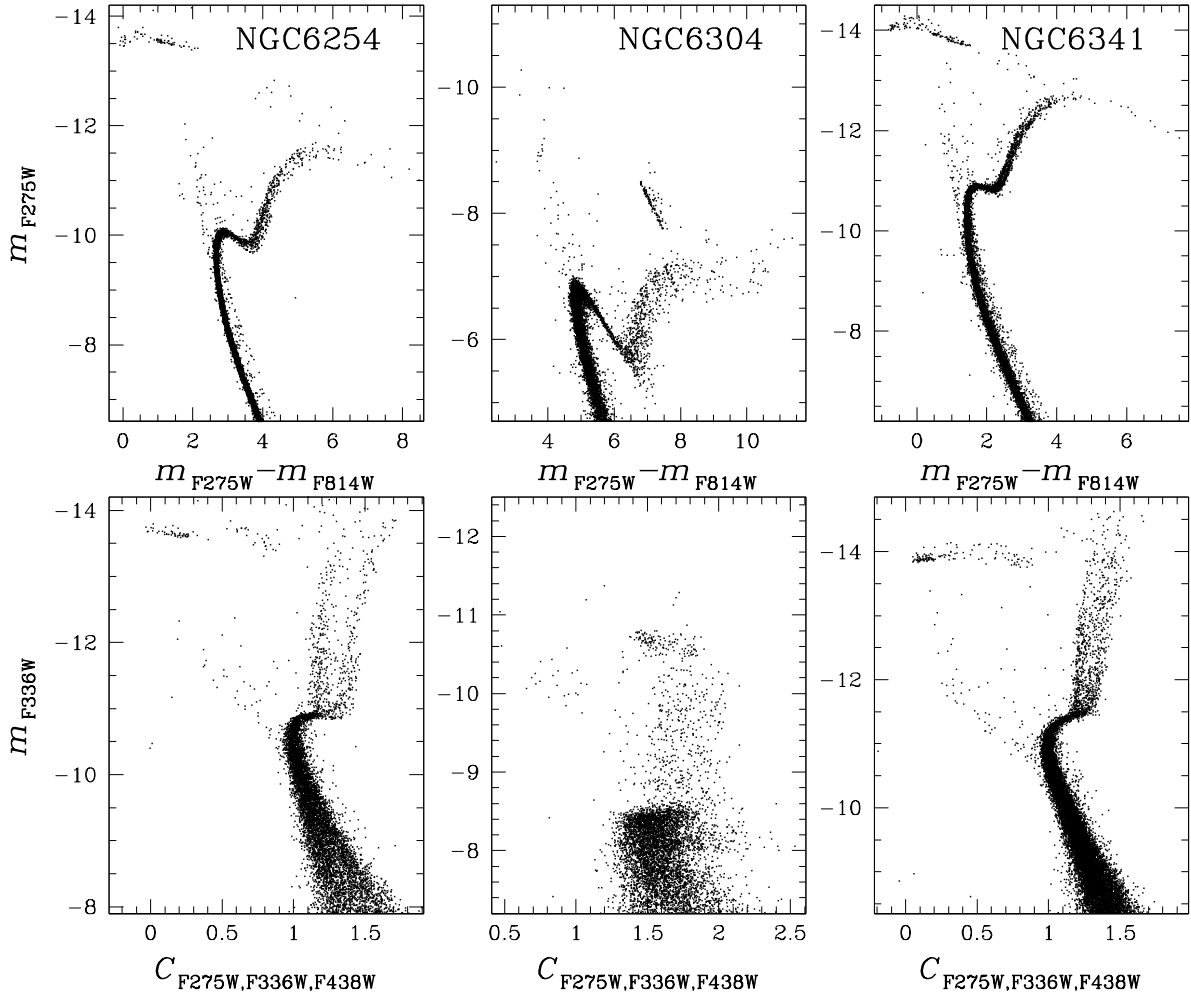}
\caption{As in Fig.~\ref{cmd1}, for NGC 6254, NGC 6304, and NGC 6341.}
\label{cmd10}
\end{figure}

\begin{figure}
\epsscale{1.00}
\plotone{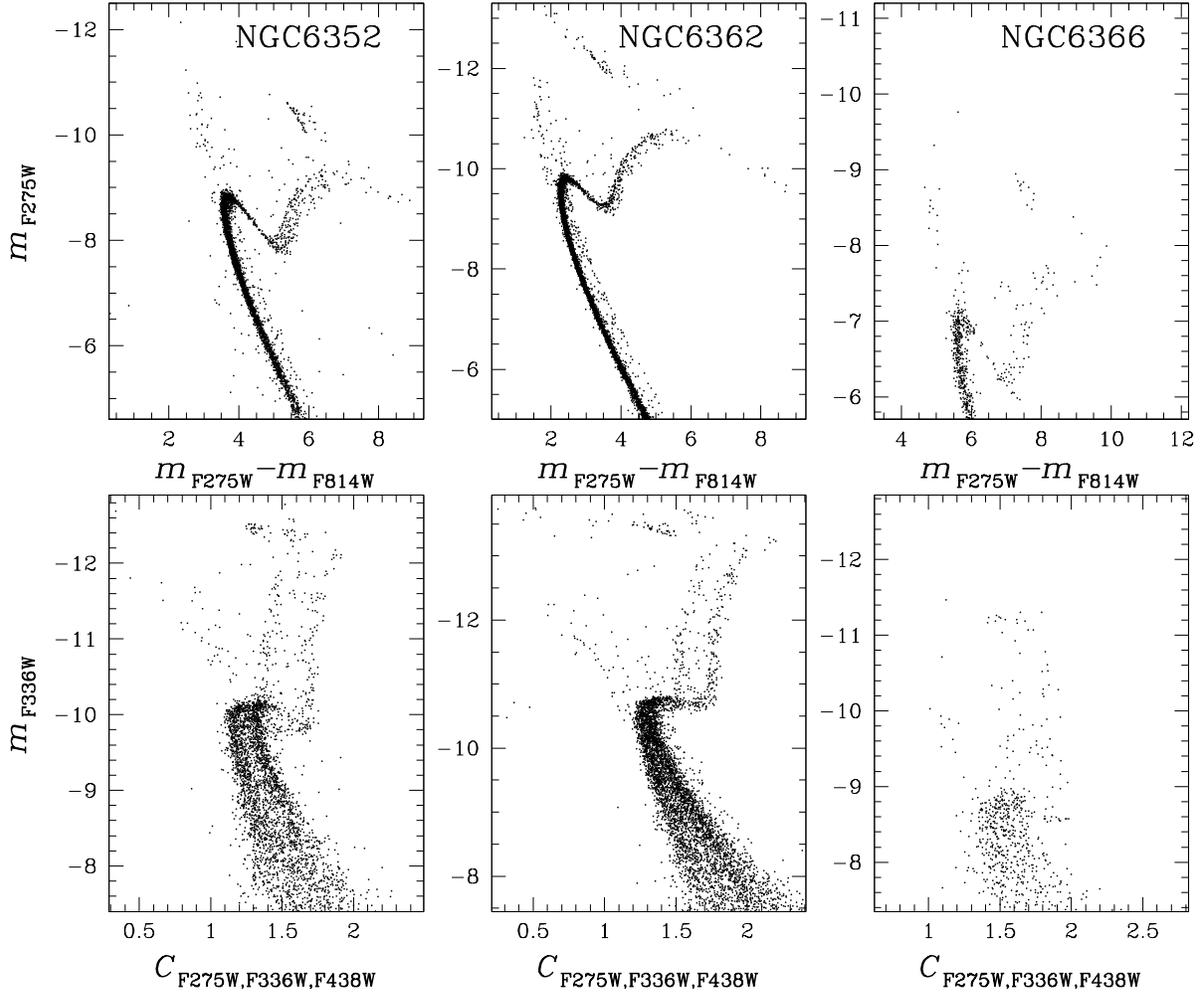}
\caption{As in Fig.~\ref{cmd1}, for NGC 6352, NGC 6362, and NGC 6366.}
\label{cmd11}
\end{figure}

\begin{figure}
\epsscale{1.00}
\plotone{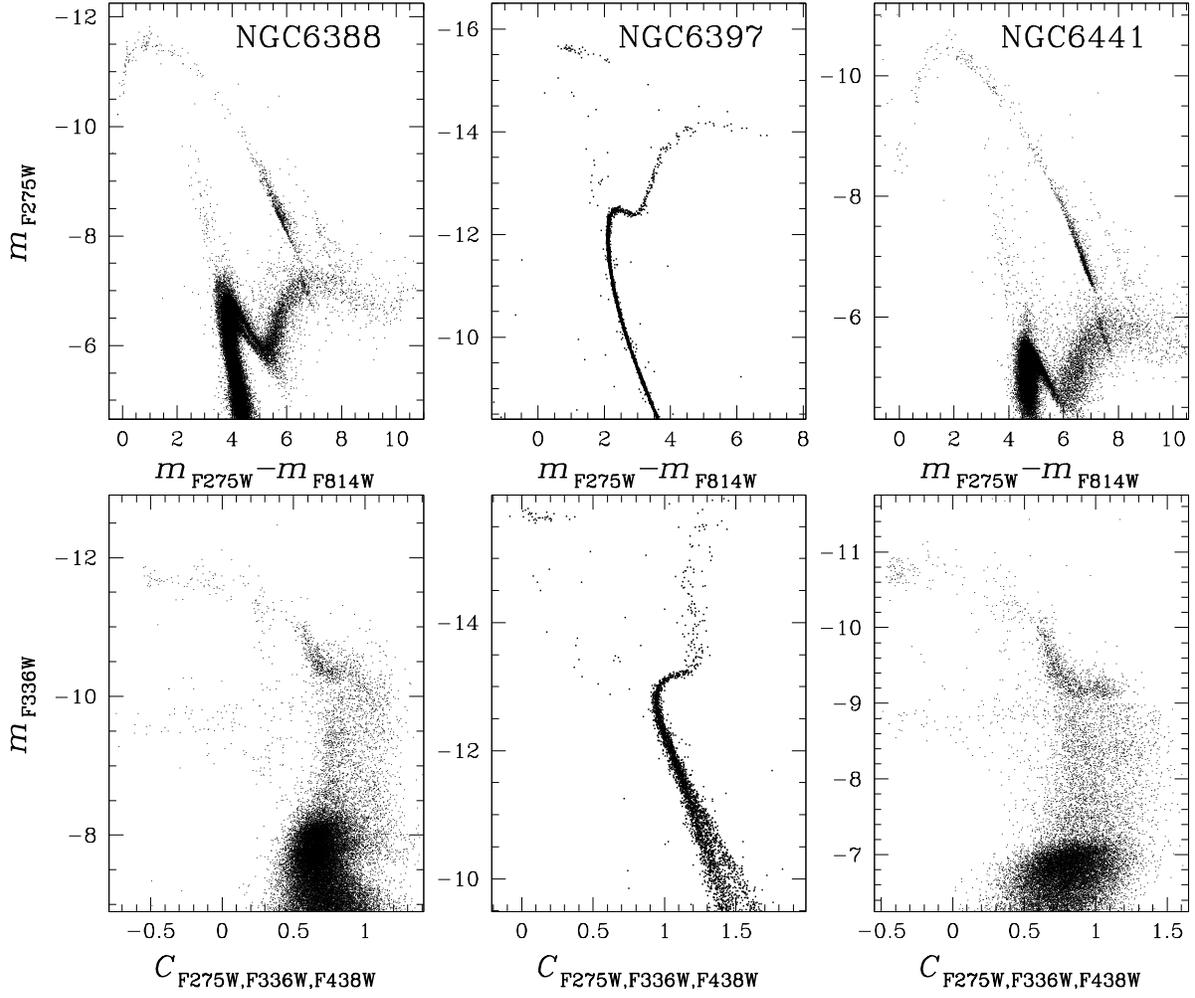}
\caption{As in Fig.~\ref{cmd1}, for NGC 6388, NGC 6397, and NGC 6441.}
\label{cmd12}
\end{figure}

\begin{figure}
\epsscale{1.00}
\plotone{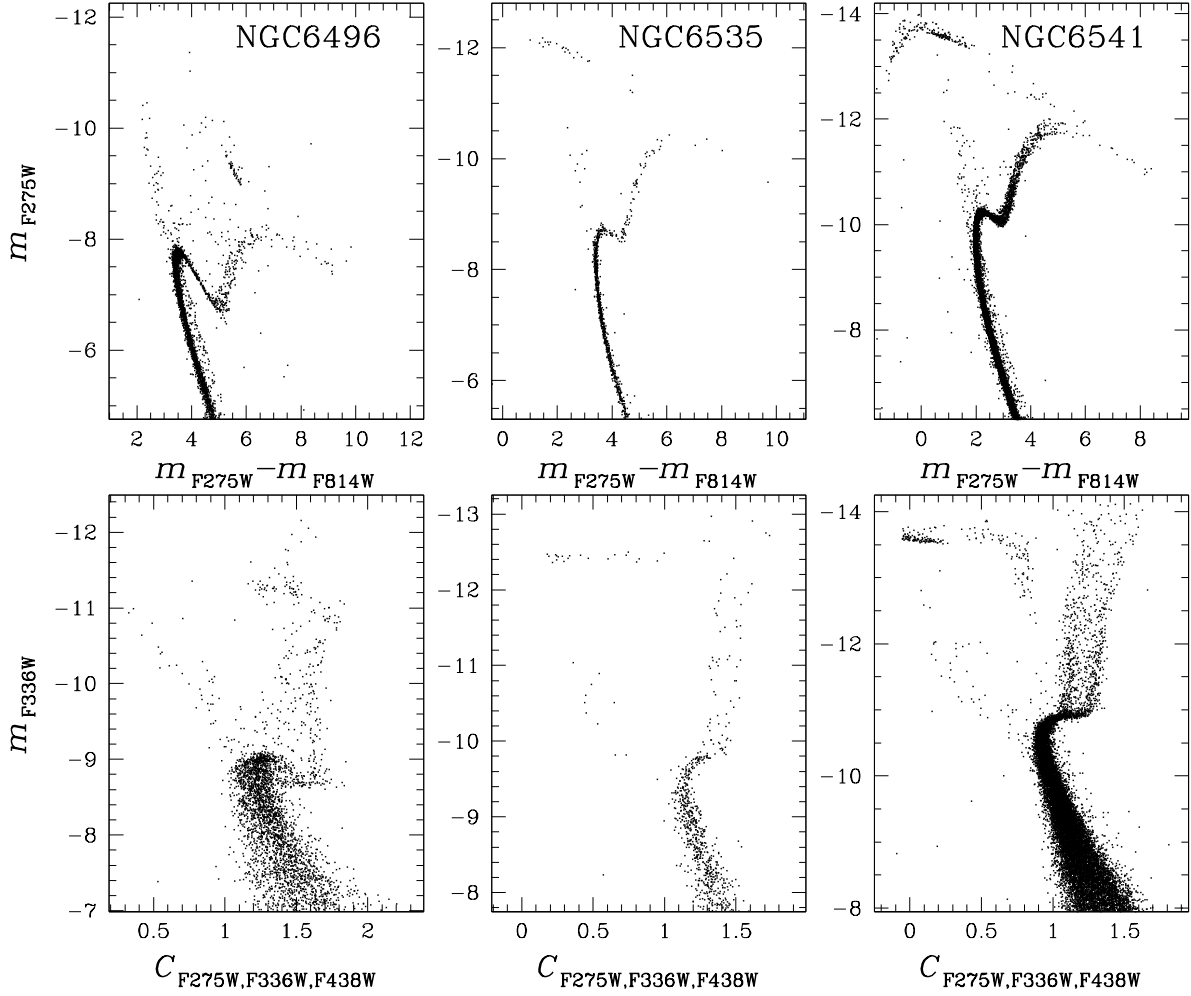}
\caption{As in Fig.~\ref{cmd1}, for NGC 6496, NGC 6535, and NGC 6541.}
\label{cmd13}
\end{figure}

\begin{figure}
\epsscale{1.00}
\plotone{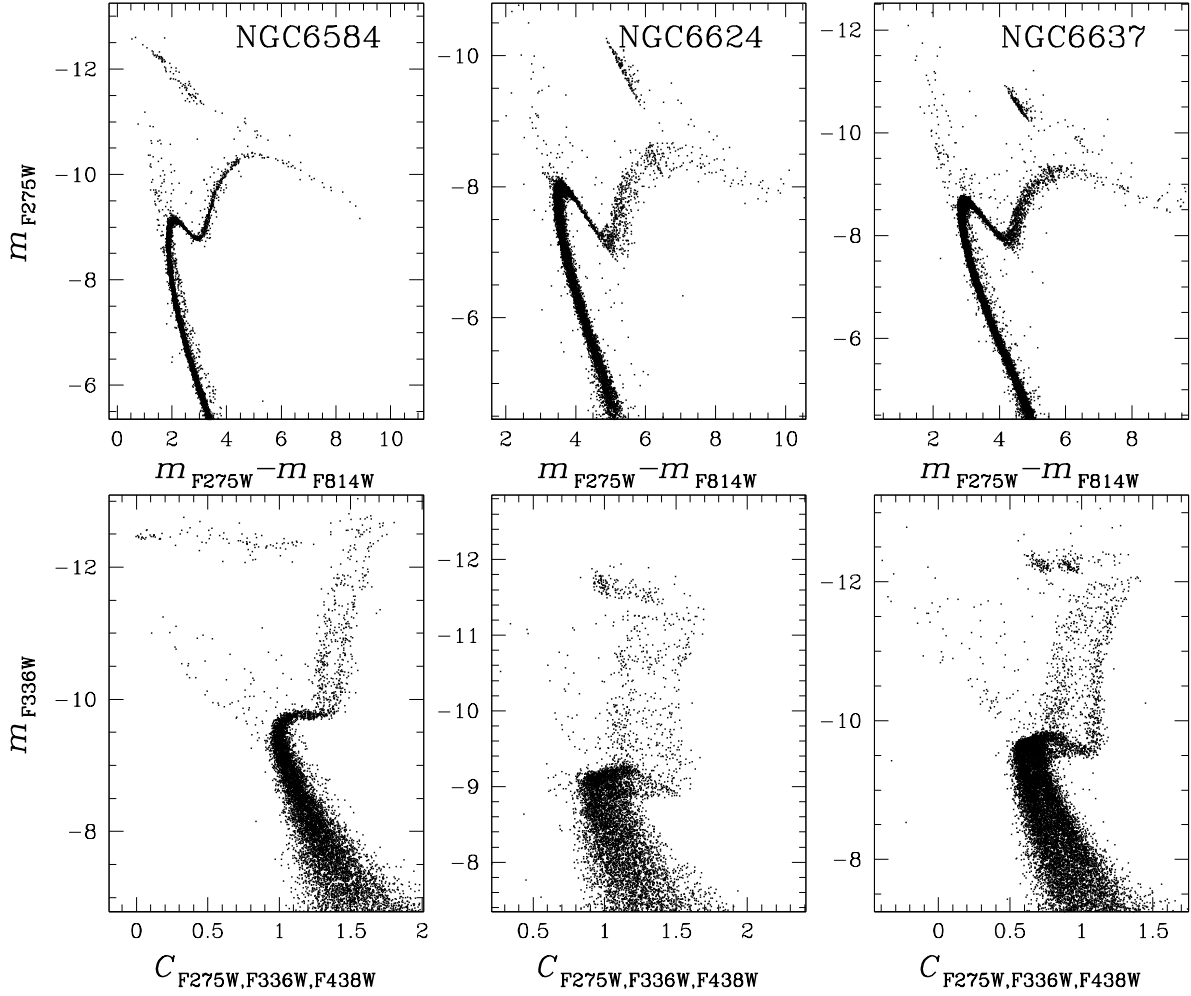}
\caption{As in Fig.~\ref{cmd1}, for NGC 6584, NGC 6624, and NGC 6637.}
\label{cmd14}
\end{figure}

\begin{figure}
\epsscale{1.00}
\plotone{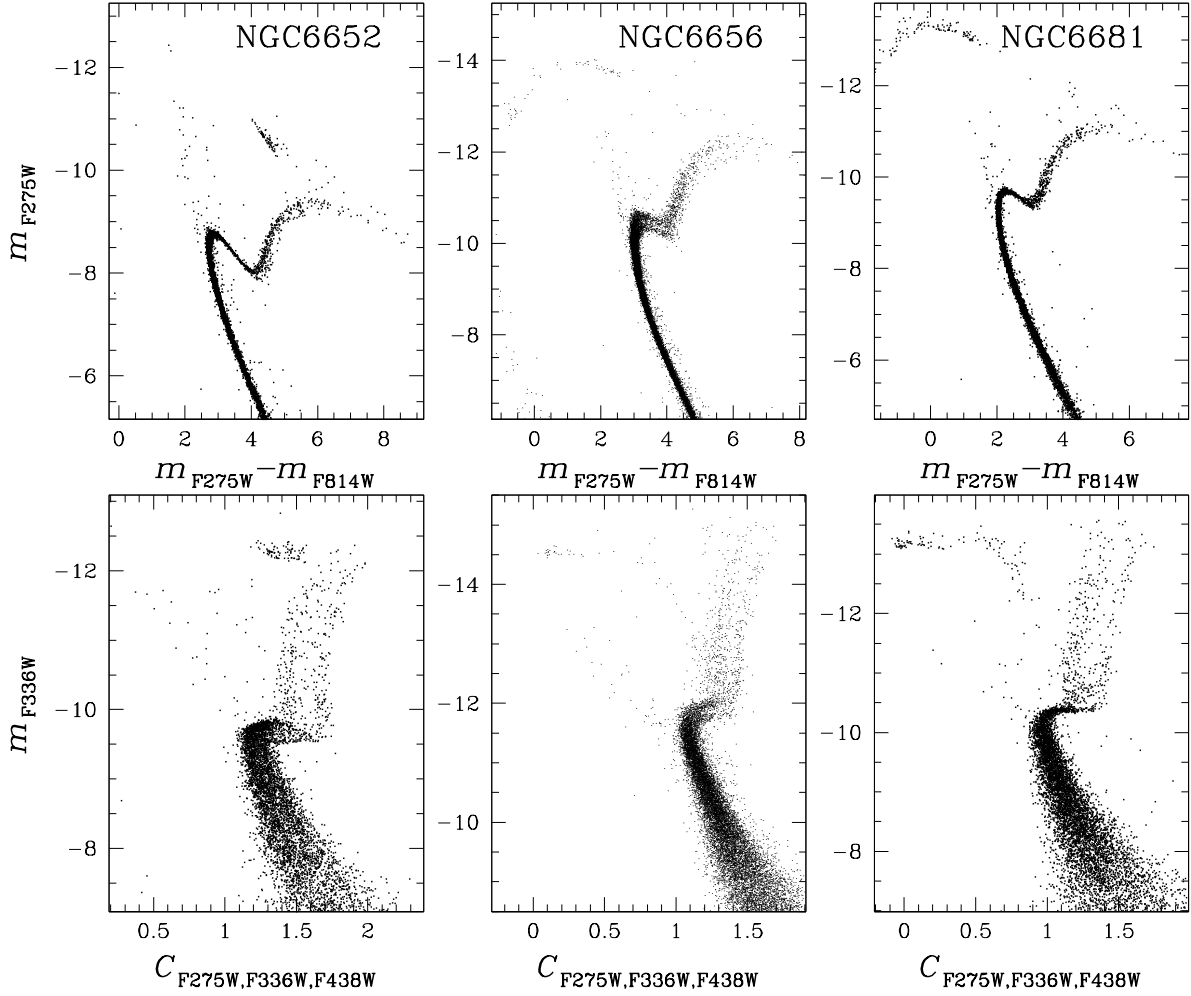}
\caption{As in Fig.~\ref{cmd1}, for NGC 6652, NGC 6656, and NGC 6681.}
\label{cmd15}
\end{figure}

\begin{figure}
\epsscale{1.00}
\plotone{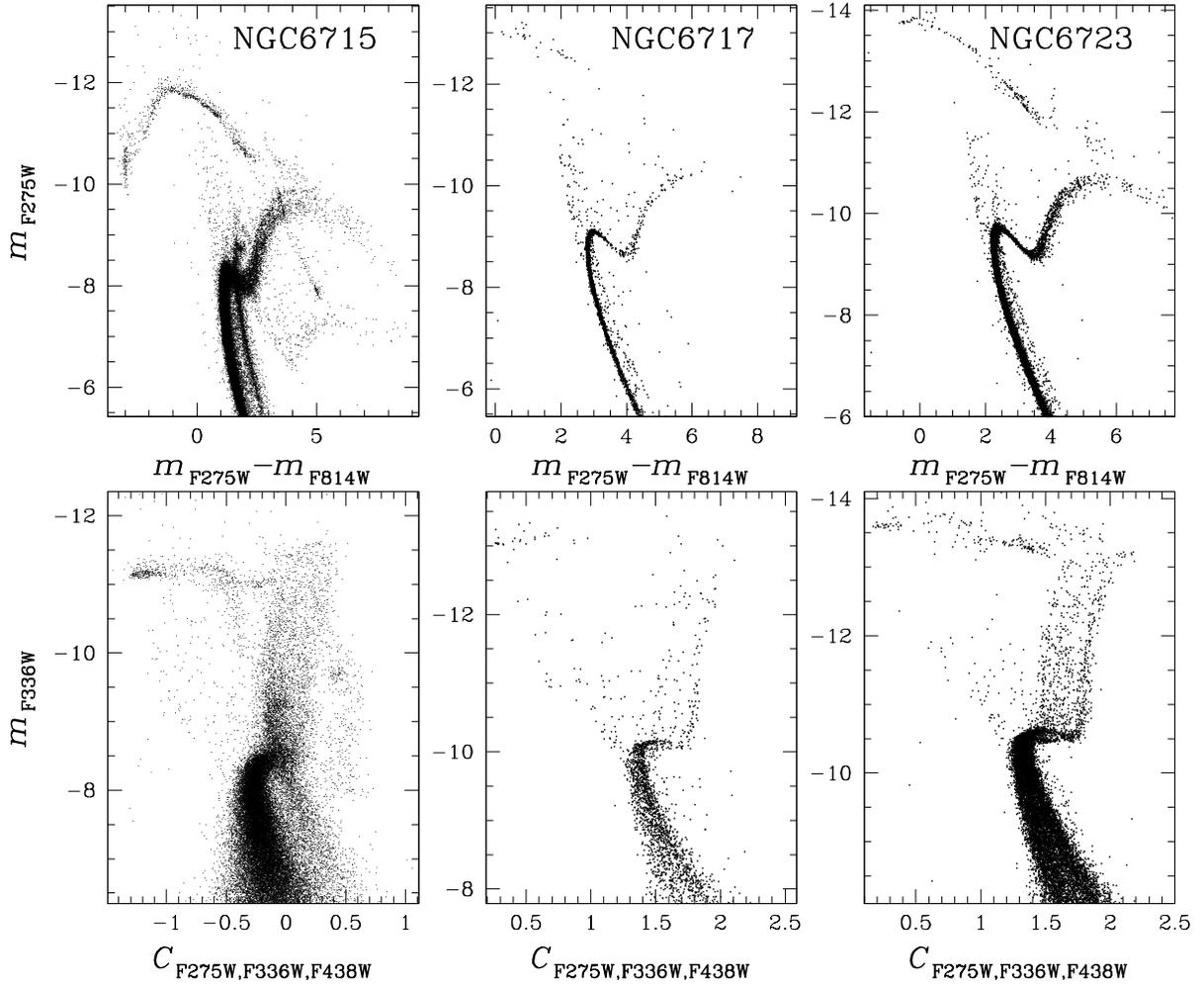}
\caption{As in Fig.~\ref{cmd1}, for NGC 6715, NGC 6717, and NGC 6723.}
\label{cmd16}
\end{figure}

\begin{figure}
\epsscale{1.00}
\plotone{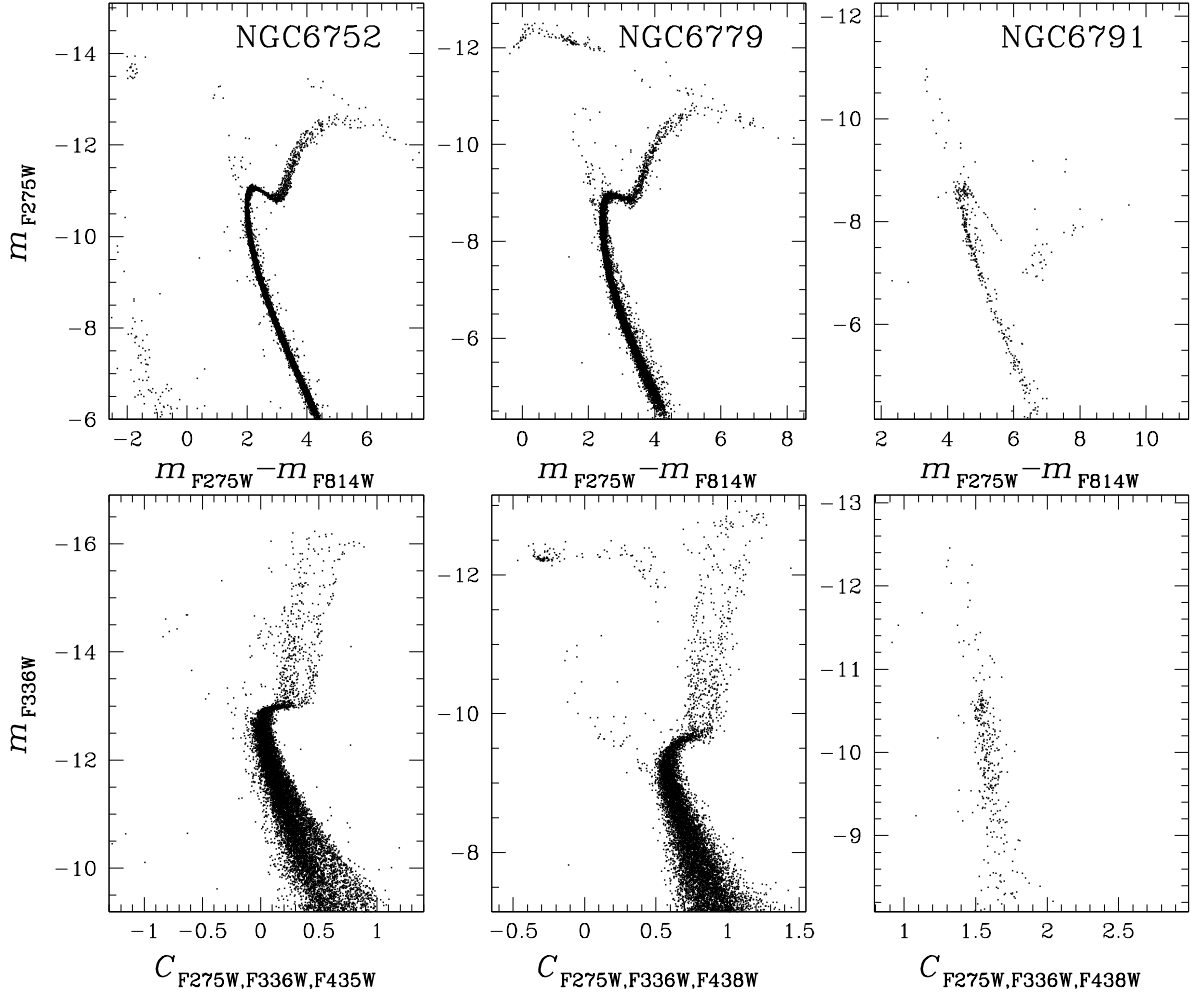}
\caption{As in Fig.~\ref{cmd1}, for NGC 6752, NGC 6779, and NGC 6791.}
\label{cmd17}
\end{figure}

\begin{figure}
\epsscale{1.00}
\plotone{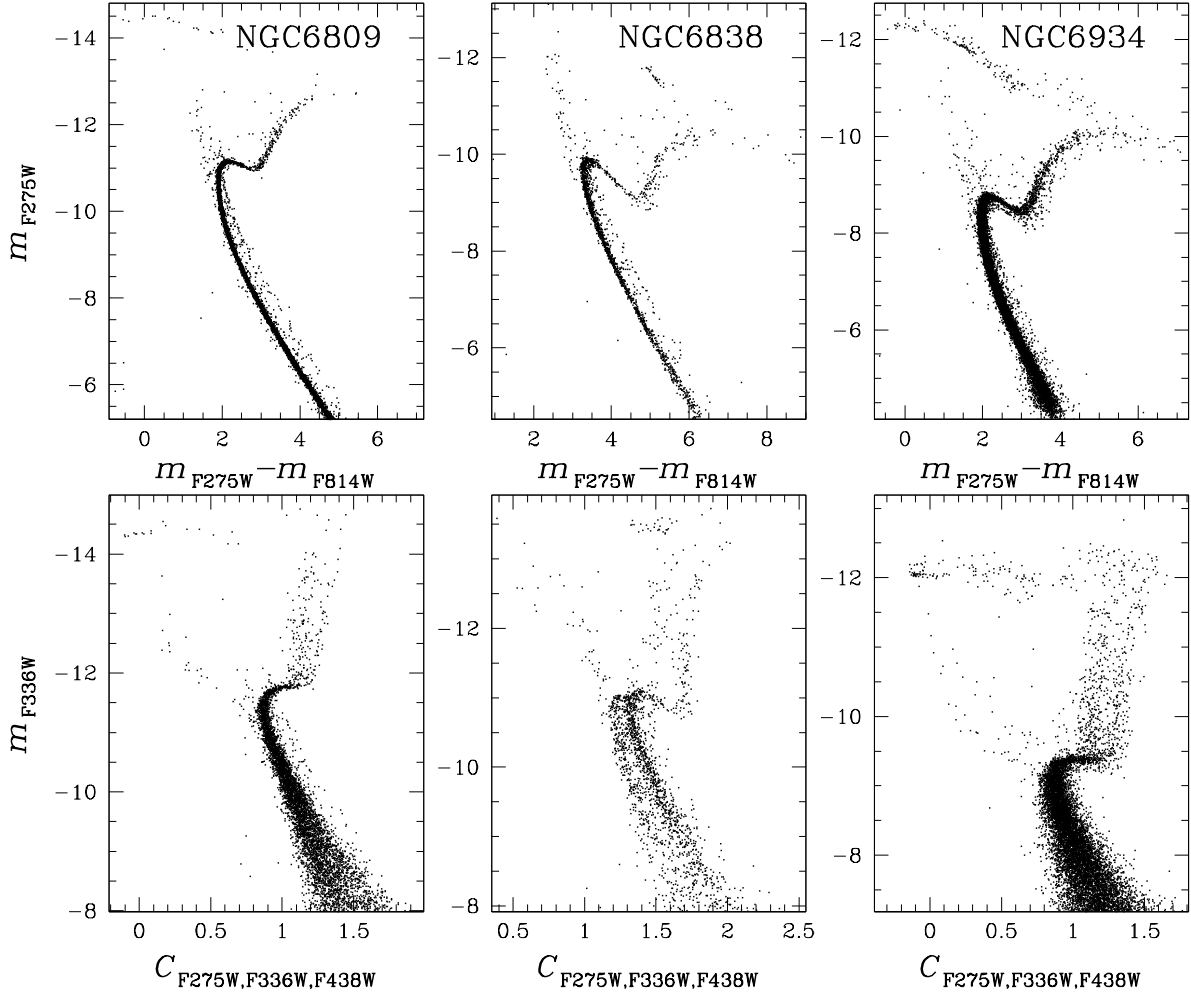}
\caption{As in Fig.~\ref{cmd1}, for NGC 6809, NGC 6838, and NGC 6934.}
\label{cmd18}
\end{figure}

\begin{figure}
\epsscale{1.00}
\plotone{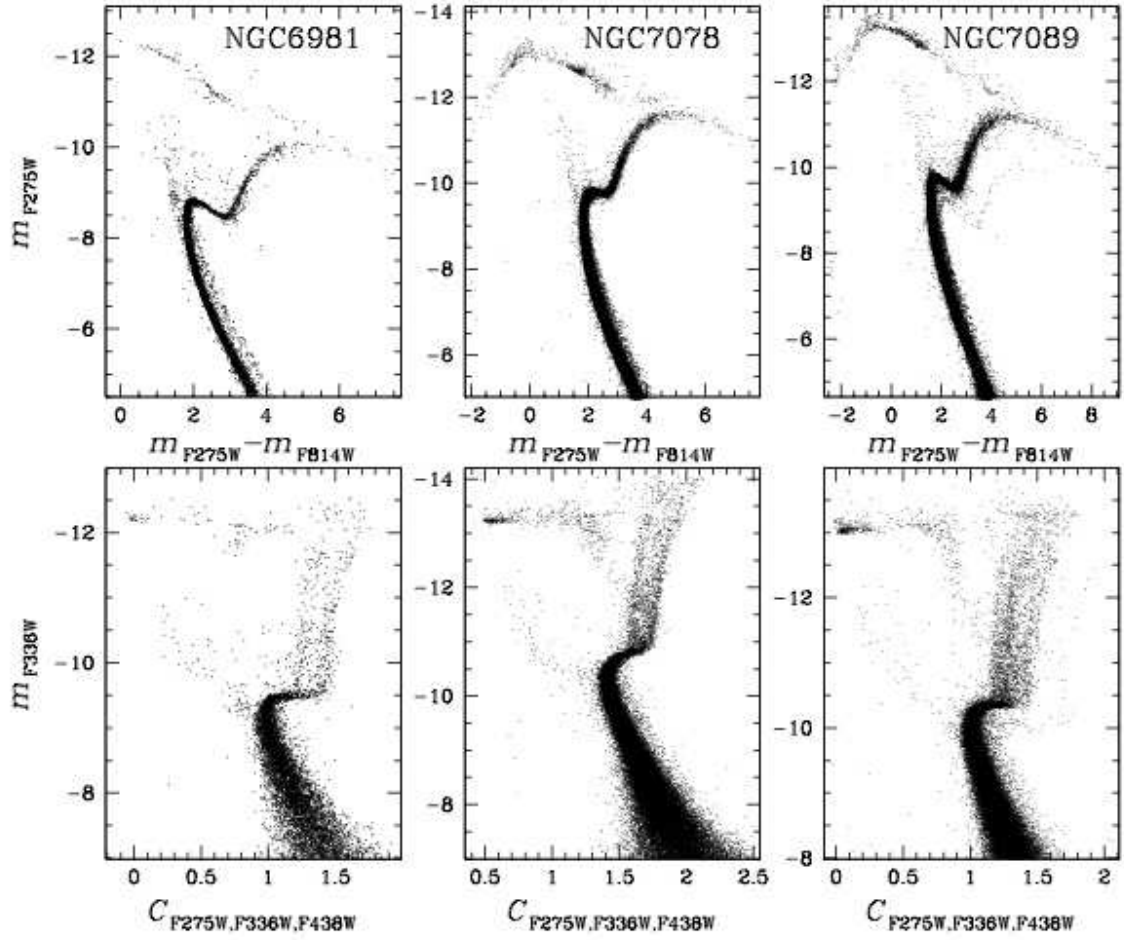}
\caption {As in Fig.~\ref{cmd1}, for NGC 6981, NGC 7078, NGC 7089.}
\label{cmd19}
\end{figure}

\begin{figure}
\epsscale{1.00}
\plotone{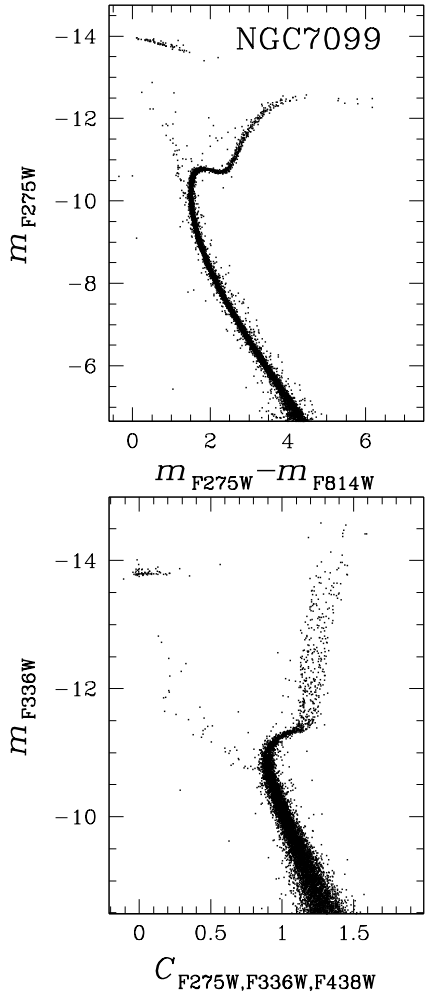}
\caption {As in Fig.~\ref{cmd1}, for NGC 7099.}
\label{cmd20}
\end{figure}

In Figs.~\ref{cmd1}-\ref{cmd20} we present the F275W vs.\ $C_{\rm
  F275W,F336W,F438W}$ and the F275W vs.\ F275W $-$ F814W diagrams.
Magnitudes are instrumental magnitudes.  Instrumental magnitudes are
defined as the ${\rm -2.5*Log_{10}}$\ of the sum of the detected
photo-electrons within a circular aperture of 10 pixels, in the
reference exposures of each filter, chosen as the one with the longest
exposure time.  A careful calibration will need appropriate aperture
correction tailored to the adopted PSFs, which differs frame by frame.
We will eventually do so, but, at the moment, in order to avoid future
confusion, we prefer to show the CMDs using the instrumental
magnitudes as above defined.  The figures show the diagrams for the
GO-13297 clusters observed (even partially, i.e., not all orbits
collected) by August, 2014.  (The figures also include the CMDs from
the two pilot projects GO-12605 and GO12311. More specifically, F275W,
F336W, F438W data for NGC 288, NGC 362, NGC 5272, NGC 6093, NGC 6205,
and NGC 7078 come from GO-12605. F275W photometry for NGC 104, NGC
1851, NGC 6121, NGC 6656, and NGC 6752 is from GO-12311. Additional
F275W and F336W data for NGC 104 and NGC 6752 come from HST archive
(see Milone et al.\ 2012a,b for details).  The F814W photometry comes
from Anderson et al.\ (2008) and is based on GO-10775 data.  F814W
photometry for NGC 6791 comes from GO10265 archive data.  For
completeness, we also added the CMD of $\omega$~Cen from Bellini et
al.\ (2010).

The CMDs are based on a preliminary selection of the best measured
stars.  There are margins for an improved photometry, and a more
complete star counts. A detailed analysis of the individual CMDs is
beyond the purposes of this paper and will be deferred to future
papers.  Still it seems appropriate to emphasize that all GCs in this
sample show multiple sequences or at least broadened sequences.  This
confirms the idea that the presence of multiple stellar populations in
GCs is a widespread phenomenon, with single-stellar-population GCs
being an exception (if, indeed, single-stellar-population GCs exist at
all).

\section{GO-13297 Project Legacy.}
\label{legacy}

The central theme of the GO-13297 project is to explore as fully as
possible the MP phenomenon in GCs, which is clearly the key to
understanding star formation and feedback in the cluster environment.
However, our database will become a gold mine for a large number of
other projects and will be a general resource for decades to come.  In
the following, we briefly describe the research projects our group
intends to carry out using the GO-13297 astrometric and photometric
database.

{\it The Identification of Multiple Stellar Populations.}  This is of
course our main goal.  We will search for MPs in all evolutionary
branches, in all clusters, following the procedure described in Milone
et al.\ (2013) for NGC 6752, and in Milone et al.\ (2012a,b) for 47Tuc
and NGC 6397.  The different populations will be chosen via different
combinations of the photometry from the five bandpasses.  We will do
this by selecting individual stars belonging to different photometric
sequences and calculating all their colors from combining the various
passbands.  In the final catalog each star will be assigned to a
particular sub-population, in order to enable spectroscopic follow-up
aimed at a precise chemical tagging of each sub-population.  The
catalogs and finding charts resulting from this effort will allow the
entire community to contribute to this critical aspect of MP studies.

{\it Modeling multiple stellar populations.}  The peculiarities
characterizing the chemical properties of the various sub-populations
hosted by Galactic GCs require the computation of extended sets of
models for low-mass stars properly accounting for these chemical
anomalies. This is a crucial step in order to investigate and trace
the distribution of the distinct sub-populations in the various
evolutionary sequences in the CMD from the faintest portion of the MS
(Milone et al.\ 2012a,b) to the HB (D'Antona \& Caloi 2008,
Dalessandro et al.\ 2011, and references therein).  The photometric
appearance of multiple populations strongly depends on the adopted
photometric bands (Sbordone et al.\ 2011, Cassisi et al.\ 2013). In
particular, since light-element anti-correlation strongly affects the
stellar spectra at wavelengths shorter than 400nm, when using UV and
far-UV passbands it is crucial to use appropriate color-$T_{\rm eff}$
relations properly accounting for the various chemical anomalies.
Basically, we will follow the method described in Milone et al.\ (2013
and references therein).

{\it Measurement of He content from multiband photometry.}
Multi-wavelength photometry provides unique information on the
chemical composition of large samples of stars belonging to the
different stellar populations hosted in GCs.  As described in
Section~\ref{magicUVs}, the combination of filters centered at
different wavelengths allows us to discriminate between various
effects, like differences {\bf in} He, C, N, O, Na content.
Exploiting the sensitivity of various colors to different properties
of stellar populations, and combining the photometric information with
grids of synthetic spectra constructed with chemical composition
constrained from spectroscopic observations will allow us to better
isolate the specific role of the helium abundance.  This technique has
been successfully applied to a few GCs, and differences in helium
among different stellar populations have been measured in 47 Tuc, NGC
6397, NGC 288 and NGC 6752 (Milone et al.\ 2013, and references
therein).  We plan to extend these measurements to the whole sample of
clusters observed for this survey.  The knowledge of the He variations
within a GC will be used to the extend investigation on how this
observable impact on the second parameter on HB morphology (see Milone
et al.\ 2014b).

{\it Cluster Structural parameters from star counts. } The appropriate
combination of the UV-survey data and complementary wide-field
observations will allow us to construct a new generation of GC radial
star density profiles based on resolved star counts (see Ferraro et
al.\ 2003, Miocchi et al.\ 2013). This is indeed the most robust and
reliable way to derive cluster structural parameters, since surface
brightness can be biased by the presence of sparse, bright stars. This
data base will allow a complete re-characterization of the structural
and dynamical properties of GCs.

{\it Internal Proper Motion Analysis. } Bellini et al.\ (2014)
recently compiled the first large sample of GC proper-motion (PM)
catalogs, using heterogeneous data from the \textit{HST} Archive (see
also Bellini et al.\ 2013b). This has shown that it is possible to
reach the level of precision and accuracy required to study in detail
the internal PM kinematics of GCs.  Among other things, this will
allow us to measure separately the kinematical properties of each of
the different stellar populations photometrically identified (as in
e.g., Anderson \& van der Marel 2010; Richer et al. 2013).  Each GC
observed as part of our GO-13297 program was also observed in 2006 by
the GO-10775 Treasury program. Moreover, a good fraction of these GCs
was also observed by other \textit{HST} programs (in particular using
the state-of-the-art detectors that have been available on
\textit{HST} since 2002). This will allow us to measure exquisite PMs
with time baselines between 7 and 14 years. This will extend the work
of Bellini et al.\ (2014) to a larger number of clusters, higher PM
accuracy, and bluer stars. Among other things, this will allow us to
measure separately the kinematical properties of each the different
stellar populations photometrically identified.  Any kinematical
differences thus identified would strongly constrain formation
mechanisms for the multiple population phenomenon.

{\it Internal Dynamical Model and Central Intermediate-Mass Black
  Holes. } The new PM catalogs to be derived from our data will allow
several new state-of-the-art studies of GC dynamics. For example, we
will obtain direct measurements of the radial velocity dispersion
profiles, the velocity dispersion anisotropy, and possible rotation of
many GCs. Dynamical modeling will constrain the mass profiles of the
GCs, including the possible presence of intermediate-mass ($\approx
10^3$ -- $10^4 M_\odot$) black holes (van der Marel \& Anderson
2010). The existence of such black holes continues to be debated in
the literature (e.g., Noyola et al.\ 2010; Lutzgendorf et al.\ 2013;
Lanzoni et al.\ 2013), but this can be tightly constrained using PMs
measured in the central arcseconds of the cluster. We will also be
able to measure the velocity dispersion as a function of main sequence
mass, and this will allow us to establish the extent to which GCs are
in energy equipartition (Trenti \& van der Marel 2013).

{\it Absolute proper motions based on an extragalactic reference
  frame. } Where possible, in our observing strategy we chose an
orientation that allowed the parallel fields to overlap previous
observations, so that we can also measure proper motions in the outer
fields. At a minimum, we can use the WFPC2 parallels taken during the
Sarajedini program, but many clusters have even better archival data.
The outer-field observations are generally deep enough to allow us to
measure absolute proper motions for the clusters with respect to
background galaxies.  The central fields will be observed in UV/blue
filters, so, although we will explore them, we do not expect to find
many background galaxies bright enough to be used as absolute points
of reference.

Absolute proper motions will be combined with radial velocities from
the literature or from measurement of archival spectra, to derive the
space velocities of the clusters; the position-velocity phase space
will then be examined for possible groupings of clusters with similar
multiple stellar population properties.

{\it Parallel fields and ground-based ancillary data for the radial
  distributions of multiple stellar populations. } Astro-photometric
catalogs and atlases in each of the filters will be electronically
provided, also for the ACS parallel fields. When possible,
proper-motion membership, and a sub-population flag will also be
provided.  Ground-based wide-field multiband photometry is already
available for a large number of the target clusters.  Merging
ground-based with the new {\it HST} data will help connect stellar
populations in the cluster centers with those in the external regions
by using similar color combinations (${\it C}_{\rm U,B,I}$ is
analogous to $C_{\rm F275W,F336W,F438W}$), and therefore investigate
their radial distribution over the whole extent of the cluster.
Moreover, identifying stars with available high-resolution spectra
will allow us to connect photometric and spectroscopic properties of
the different stellar populations.  Observations with ground-based
facilities (typically 8-10m size telescope) have already been
acquired, and additional observations requested for a subsample of the
GO-13297 clusters

{\it Modeling the radial distribution of multiple stellar
  populations. } Theoretical models and simulations predict that 2G
stars should initially be more concentrated than 1G stars (see, e.g.,
D'Ercole et al.\ 2008, Bekki 2011, Bastian et al.\ 2013b, and
references therein), and that many clusters should still preserve, at
least in part, some memory of the initial differences in the 1G/2G
spatial distribution (Vesperini et al.\ 2013; for observational
studies that show radial gradients in the fraction of 2G stars see,
e.g., Sollima et al.\ 2007, Bellini et al.\ 2009, Lardo et al.\ 2011,
Milone et al.\ 2012).  We will carry out a systematic observational
and theoretical study aimed at addressing a number of key questions
concerning this issue.  By studying the presence and extent of radial
gradients in clusters with different dynamical properties and at
different phases of their dynamical evolution we will be able a) to
further test the general validity of models predicting the initial
spatial segregation of the different stellar populations, b) to make
progress in the observational identification of an evolutionary
sequence in the spatial mixing of 1G and 2G stars, and c) to shed
light on the relationship between the global 2G fraction and its local
value measured at different distances from the cluster center.

{\it Distances.}  Knowing the distances to Milky Way GCs is
fundamental for several astrophysical fields, including stellar
evolution, cluster dynamics, and the cosmological distance ladder. To
date, GC distances have generally been obtained mostly through
isochrone-fitting techniques and the RR Lyrae variables.  By comparing
high-precision \textit{HST} PMs (at the level of mas/yr) with existing
line-of-sight velocity data (in km/s), it becomes possible to obtain
entirely independent kinematical distance estimates (either from the
assumption of velocity isotropy or from detailed dynamical
models). Such a study is currently in progress (Watkins et al., in
prep.) using the PM catalogs of Bellini et al.\ (2014). With the new
PM catalogs to be derived from the GO-13297 data, we plan to extend
this work to a larger number of clusters, and higher distance
accuracies.  These distances will be completely independent from, and
in some ways complementary to GAIA measurements.

{\it RGB bump. } We plan to measure the RGB bump brightness and its
extension in magnitude of the various sub-populations hosted by each
cluster in our sample.  Comparison with models will allow us to
estimate He-abundance differences among them.  These differences can
be compared with the He abundance measurements obtained with other
methods (see previous sections).  At the same time, the relative
brightness of the RGB bump with respect other CMD features, such as
the HB luminosity and the MSTO, will provide useful tests for stellar
evolution models (e.g., Cassisi et al.\ 2011).

{\it HB morphology. } Because optical colors become degenerate at high
temperature, our use of UV bands will enable an accurate
characterization of the full HB morphology, including the extreme
horizontal branch (EHB).  The HB morphology will yield additional
insight into any sub-populations enhanced in helium.  At higher $Y$,
the HB generally extends to hotter temperature, and also exhibits
significant variations in luminosity, with higher luminosity on the
blue HB and lower luminosity on the EHB.  It will also be possible to
establish to what extent the long-standing issue of the ``second
parameter'' can be accounted for by helium-enriched second
generations.  Some of us have already shown that the multi-modal HB of
NGC 2808 (D’Antona \& Caloi 2004, Dalessandro et al.\ 2011), as well
as of the metal-rich GCs NGC 6388 and NGC 6441 (Busso et al.\ 2007,
Caloi \& D’Antona 2007, Bellini et al.\ 2013a) is consistent with
multiple stellar populations with different helium abundance.  This
connection is supported by recent studies, based on spectroscopy of HB
stars, which have shown that distinct stellar populations, with
different content of light elements and helium, occupy different
regions along the HB (Marino et al.\ 2011, 2013, 2014; Villanova et
al.\ 2012; Gratton et al.\ 2011, 2012, 2013).  Additional examples of
the possible effects of helium enhancement on the HB morphology for
many other clusters can be found in Brown et al.\ (2010) and
Dalessandro et al.\ (2013a).  With the new GO-13297 we can extend the
investigation of Milone et al.\ (2014b) on the nature of the GC global
and local parameters and their effect on the HB morphology.

{\it Modeling HBs. } The HB morphology is very sensitive to the main
parameters of the multiple populations, and in addition, to mass loss
on the red giant branch. We need to build synthetic HBs, based on the
computation of large grids of models, to check and quantify the role
of all the parameters. New grids are being computed (Tailo et al.\
2014, in preparation), including models populating the highest
$T\_{eff}$'s of the HB. These structures are the outcome of ``late
helium core flash" ignition (D'Cruz et al.\ 1996, Brown et al.\
2001). A crucial outcome of the comparison with observations will be
to verify whether the very hot HB stars result from a high-helium
population. As a byproduct of the population synthesis for the HB, we
also predict how large a population of helium white dwarfs is expected
in clusters hosting extreme HB stars, as found by Bellini et al.\ (2013c)
in $\omega$~Cen.

{\it Analysis of RGB+HB+AGB star counts (searching for AGB manqu\'e
  objects). } Spectroscopic studies have revealed that the AGBs of
some GCs host only CN-weak/Na-poor stars, at odds with what is
observed along the RGB, where both CN-weak and CN-strong stars are
present (e.g., Norris et al.\ 1981, Pilachowski et al.\ 1996, Campbell
et al.\ 2010, 2013).  This has been interpreted as the presence of
multiple stellar populations with different nitrogen and helium
abundances.  The helium-rich/nitrogen-rich stars would populate the
hottest part of the HB, and then evolve without ascending the AGB,
thus making the AGB populated by CN-weak stars only (Norris et al.\
1981).  We plan to analyze star counts of RGB, HB, and AGB stars to
further investigate the connection between other stellar populations
and AGB-manqu\'e objects.

{\it Globular Cluster Ages. }  Relative ages of 64 globular clusters
were measured in the ACS Treasury Program of Galactic Globular
Clusters (Marin-Franch et al.\ 2009).  Single stellar populations were
assumed, however, and no dependence on chemical abundances was taken
into account; these limitations were due to lack of
composition-dependent stellar-evolution libraries and to having only
two-color photometry.  The GO-13297 data set now adds three more
UV/optical bands, and the resulting set of five photometric bands is
much better able to disentangle multiple stellar populations, which
are also distinguished on the basis of their CNO content.  We are also
computing a new stellar-evolution library, including varying chemical
compositions, which will enable us to interpret our new photometry in
terms of MPs in GCs.  In particular the new models include variation
of CNO content, which has a strong impact on age estimates (Cassisi et
al.\ 2008, Ventura et al.\ 2009).  The final product will be a new
estimate of relative ages for the different populations, This will be
done with a precision that has never been attained before.

{\it Binaries in multiple stellar populations. }  An accurate study of
the CMD can provide information on the population of binary systems in
GCs, such as the fraction of MS--MS binaries, the distribution of mass
ratios, and the radial distribution (see, e.g., Romani \& Weinberg
1991, Sollima et al.\ 2007, Milone et al.\ 2012c).  These are
fundamental ingredients for understanding the dynamical evolution of
GCs and their population of exotic stellar objects, such as BSSs,
cataclysmic variables, millisecond pulsars, and low-mass X-ray
binaries.  We plan to exploit the large number of CMDs that can be
obtained from our data set to refine our previous investigations on
binary fraction, binary mass distribution, and binary radial
distribution (Milone et al.\ 2012c), which was based on F606W and
F814W photometry only.  Moreover, in clusters with multiple MSs, we
will estimate the fraction of binaries in the different stellar
populations.

{\it Blue Straggler Stars. }  Blue straggler stars are crucial probes
of the internal dynamics of globular clusters (e.g., Ferraro et al.\
2012). Since they are hotter than 6000--7000 K, they are best
detectable at UV wavelengths, so that we can use the UV survey data to
obtain complete samples of BSSs in the central regions of GCs, with a
significant improvement in completeness over that of previous efforts
(e.g., Piotto et al.\ 2004).  Detailed star counts and complete
luminosity functions will be obtained, possible sub-structure unveiled
(such as the double sequences in M30 and NGC 362; Ferraro et al.\
2009b, Dalessandro et al.\ 2013b), and extended radial distributions
will be derived from combination with wider-field ground-based
observations.

{\it Stellar exotica. }  The UV survey will be used to search for
exotic objects that came from the evolution of binary systems in the
cores of GCs.  Among these, of particular interest is the
identification of the companion stars in binary millisecond pulsars
(Ferraro et al.\ 2001, 2003b; Pallanca et al.\ 2010), since it would
allow us to quantify the occurrence of dynamical interactions and to
constrain the mass of neutron stars, with invaluable consequences for
the equation of state of matter at equilibrium densities that are near
nuclear.

{\it Cross correlation of X-ray sources with HST UV/optical data. }
More generally, studies of the nature, origin, and evolution of binary
stars and their progeny will be aided by cross-correlation of the
GO-13297 data with imaging at X-ray wavelengths.  As repositories of
energy, binaries play a central role in the dynamics of globular
clusters.  With sub-arcsecond positions provided by the Chandra X-ray
Observatory, whose archive already includes more than half of the
survey clusters, optical and/or UV counterparts can be identified for
a wide variety of systems.  These include cataclysmic variables,
quiescent low-mass X-ray binaries, millisecond pulsars, red
stragglers, and systems with active coronae (e.g., RS, CVn, and BY Dra
stars).  Proper motions measured as a part of this program will make
the additional contribution of separating binaries and stellar exotica
from field stars that occupy the same regions of the CMD.

{\it Luminosity and Mass Functions of Different Stellar Populations. }
GO-10775 allowed us to extract the stellar mass functions for 17 GCs
(Paust et al.\ 2010).  With the new data we plan to extend this
survey; this will have important implications, in particular for the
effect of the Galactic gravitational potential (Djorgovski et al.\
1993) on the GC stellar mass function.  Moreover, from CMDs that now
include UV colors, with which multiple MSs are better distinguished,
we should be able to obtain the mass functions of the different
stellar populations in the same cluster, as Milone et al.\ (2012d) did
for NGC 2808.

{\it Galactic Bulge Globular Clusters. } Multi-populations have been
detected in a few massive Galactic-bulge GCs, but so far no systematic
search for them has been undertaken.  About half a dozen clusters in
the GO-13297 target list are located in the inner bulge or at the edge
of the bulge/halo transition zone, and may be dynamically connected to
the bulge. The bulge clusters have different properties from the
classical halo clusters, and there is evidence that they formed in an
early epoch (Barbuy et al.\ 1998, Cot\'e 1999).  The metal-rich
clusters there might show a high helium abundance.\\
A few other clusters show a blue horizontal branch and a metallicity
[Fe/H] $\geq -1.0$. Therefore these GCs are either very old or else
they host at least two stellar populations.

{\it Spectral Energy Distributions of Multiple Stellar Populations. }
We will use the multi-band photometry to calculate integrated cluster
UV/optical colors by summing the flux of all measured stars in order
to obtain an integrated-color database that can be used for the study
of extragalactic GCs.  More generally, GCs are widely used as template
stellar populations for the calibration of stellar evolutionary
sequences and of the synthetic stellar populations that are
constructed with them. These are then used to interpret the light from
very distant galaxies, including the rest-frame UV. The multiband
database resulting from this Treasury program will constitute the main
benchmark for this indispensable calibration.

{\it Link between Globular Cluster UV Properties and Extragalactic
  Sources. } With these data it will be possible to calibrate
integrated cluster UV/optical colors for the diagnostics of MPs in
unresolved stellar systems, such as GCs in galaxies well beyond the
Local Group, exploring to what extent the formation of GCs has
proceeded in a similar (or different) fashion in different galactic
environments, e.g., in ellipticals vs.  spirals. Indeed, the suspicion
exists that several massive GCs in the giant elliptical M87 may harbor
helium-enriched populations as revealed by UV excess, likely due to
the presence of very hot 2G HB stars (e.g., Kaviraj et al.\ 2007).

{\it Differential reddening analysis and reddening maps. } The
interstellar medium exhibits complex structure, on scales ranging from
a few astronomical units to several kiloparsecs (e.g., Schlegel et
al.\ 1998).  We plan to determine for each cluster a high-resolution
map (typically around 10 square arcsec) of differential reddening (as
in Milone et al.\ 2012b, Sect.~3).  Correcting photometry for
differential reddening is a crucial step for identifying multiple
stellar populations in the CMD of any GC.

{\it UV Photometric standards. } To date there is no large space-based
network of UV photometric standards.  We propose to use HST UV
archival material to improve the existing absolute photometric
calibration of WFC3/UVIS in F275W and F336W bands, and to build a
network of photometric standards that will remain a fundamental
reference for years to come.  GO-13297 GCs are rather evenly
distributed in the sky and contains hundreds of bright and relatively
isolated (at least in UV images) stars that can be used to calibrate
other UV instruments, and the ground-based U-band observations.  This
is particularly valuable taking into account that after {\it HST}, no
other large space-based observatory with UV capability is expected to
fly for at least 10 years.

{\it Serendipity. } We are certain that in addition to those mentioned
above, this multi-band legacy database for Galactic GCs will offer
unlimited possibilities for serendipitous discoveries and an extremely
wide variety of other scientific investigations, and will contribute
to our understanding of the formation of our Galaxy itself.

\acknowledgements AA, SC, SH, MM, and GP recognize partial support by
the IAC (grant P301031) and the Ministry of Competitiveness and
Innovation of Spain (grant AYA2010-16717).  JA, AC, IRK, AS, and EV
acknowledge support from STScI grant GO-13297.  APM acknowledges
support by the Australian Research Council through Discovery Project
grant DP120100475.  MZ acknowledges support by Proyecto Fondecyt
Regular 1110393, by the BASAL Center for Astrophysics and Associated
Technologies PFB-06, and by Project IC120009 "Millennium Institute of
Astrophysics (MAS)" of Iniciativa Cient\'\i fica Milenio by the
Chilean Ministry of Economy, Development and Tourism.  FRF, BL and ED
acknowledge the support from the Cosmic-Lab project (web site:
http://www.cosmic-lab.eu) funded by the European Research Council,
under contract ERC-2010-AdG-267675.

\newpage

\newpage

\newpage
\clearpage

\newpage
\clearpage

\newpage
\clearpage

\newpage
\clearpage

\newpage
\clearpage

\newpage
\clearpage

\newpage
\clearpage

\newpage
\clearpage

\newpage
\clearpage

\newpage
\clearpage

\newpage
\clearpage

\newpage
\clearpage

\newpage
\clearpage

\newpage
\clearpage

\newpage
\clearpage

\newpage
\clearpage

\newpage
\clearpage

\newpage
\clearpage

\begin{landscape}

\newpage
\begin{table}
\footnotesize
\begin{tabular}{l c |c c c |c c |c c |c c c c c}
\hline
\hline
                    &           & \multicolumn{3}{|c|}{Primary Fields} & \multicolumn{2}{|c|}{Parallel Fields} & \multicolumn{2}{|c|}{GO-10775 Fields} &        &       &       &          &  \\
Cluster             & \# Orbits & Exp.\,Time  & Exp.\,Time & Exp.\,Time & Exp.\,Time & Exp.\,Time & Exp.\,Time & Exp.\,Time &  X     & Y     & Z     & $E(B-V)$ & $(m-M)_V$ \\
                    &           & F275W[s]    & F336W[s]   & F438W[s]   &F475W[s]    & F814W[s]   & F606W[s]   & F814W[s]   & [kpc]  & [kpc] & [kpc] &          &           \\
\hline
NGC\,104$^*$&     2     &$12\times348$,   &         &                & $2\times265$  &   & $4\times50$,& $4\times50$,&   -6.4   &    -2.6    &   -2.7 & 0.04 &  13.37 \\	
                    &           &    $2\times323$           &               &              &             &              &  3          & 3          &       &       &        &      &  \\ 
\hline
NGC\,288$^{**}$&     2     &$6\times401$   & $4\times350$ &$4\times41$      &   &   & $3\times10$,& $3\times10$,& 	 -8.4   &    0.0   &   -8.5 & 0.03  & 14.84 \\
                    &           &               &               &              &             &              &  $12\times130$          & $12\times150$          &       &       &        &      &  \\ 
\hline
NGC\,362$^{**}$&     2     &$6\times519$   & $4\times350$ &$4\times54$      &   &   & $4\times150$,& $4\times170$,& -5.2   &   -5.1   &   -5.8 &0.05 &  14.83 \\
                    &           &               &               &              &             &              &  10          & 10          &       &       &        &      &  \\ 
\hline
NGC\,1261           &     5     & $4\times834$,& $2\times413$,& 164, 165,    & $2\times745$,& $2\times669$,& $5\times350$,& $5\times360$,&   -8.2   &   -10.0  &  -12.5 & 0.01 & 16.09 \\ 
                    &           & $2\times855$,& $2\times415$,& 167, 168,    & 766, 770,    & 690, 697,    &      40      &     40       &       &      &        &      &  \\ 
                    &           & $2\times859$,& 419          & 170          & 829          & 753          &              &              &       &      &        &      &  \\ 
                    &           & $2\times918$ &              &              &              &              &              &              &       &      &        &      &  \\ 
\hline
NGC\,1851           &     1     &              & $4\times453$ & $2\times140$ &              &              & $5\times350$,& $5\times350$,&  -12.5   &    -8.9  &   -6.5  & 0.02 &  15.47 \\
                    &           &              &              &              &              &              &     20       & 20           &       &      &        &      &        \\
NGC\,1851$^*$&     7     &$14\times1280$  &         &                & $7\times40$  &  $7\times976$ &        &            &       &      &        &      &        \\
                        &           &                &         &                & $7\times947$ &              &           &            &       &      &        &      &        \\
\hline
NGC\,2808$^{**}$&     6     &$12\times985$   & $6\times650$ &$6\times97$  & $6\times890$,  & $6\times508$   & $5\times360$,& $5\times370$,&  -6.3   &   -9.2   &   -1.5 & 0.22 &  15.59 \\
                    &           &               &               &              &   $6\times982$   &                &  23          & 23          &       &       &        &      &  \\ 
\hline
NGC\,2298           &     4     & $4\times848$,& $4\times350$ & $2\times134$,& $2\times785$,& $2\times683$,& $5\times350$,& $5\times350$,&   -12.6   &    -9.4  &   -2.6  & 0.14 &  15.60 \\  
                    &           & $2\times980$,&              & 136,137      & 885, 887     & 815, 816     &      20      &     20       &       &      &        &      &  \\ 
                    &           & $2\times981$ &              &              &              &              &              &              &       &      &        &      &  \\ 
\hline
NGC\,3201           &     2     & $2\times754$,& $4\times310$ &  60, 68      & 685, 689     & 612, 616     &$1\times5$,   & $1\times5$,  &   -7.7   &    -4.8  &    1.1   & 0.24 &  14.20\\ 
                    &           & $2\times758$ &              &              &              &              &$4\times100$,  & $4\times100$ &       &      &        &      &  \\ 
\hline
NGC\,4590 (M\,68)   &     2     & $4\times696$ & $4\times305$ & $2\times60$  & $2\times627$ & $2\times554$ &$4\times130$, & $4\times150$,&   -4.2   &    -7.2  &    6.4 & 0.05 &  15.21 \\
                    &           &              &              &              &              &              &        12    & 12           &       &      &        &      &  \\ 
\hline
NGC\,4833           &     4     & $4\times936$,& $4\times350$ & $2\times136$,& $2\times806$,& $2\times771$,& $1\times10$, & $1\times10$, &  -4.7   &    -5.4  &   -0.5  & 0.32 &  15.08 \\ 
                    &           & $4\times895$ &              & 135,138      & $2\times840$ & $2\times730$ & $4\times150$, & $4\times170$, &       &      &        &      &  \\ 
\hline
NGC\,5024 (M\,53)   &     6     &$2\times1733$,& $3\times433$,& $3\times170$,&$2\times1451$,& $3\times370$,& $5\times340$,& $5\times340$,& -5.5   &    -1.4  &   18.0 & 0.02 &  16.32  \\
                    &           &$2\times1832$,& $3\times460$ & $2\times180$,&$2\times1550$,& $3\times375$ &  45          & 45           &       &       &        &      &  \\ 
                    &           &   1729       &              &   178        & 1447, 1548   &              &              &              &       &      &        &      &  \\ 
\hline
NGC\,5053           &     5     &$4\times829$,& $5\times415$ & $2\times164$,&$2\times740$,& $2\times664$,& $5\times340$,& $5\times350$,& -5.3   &    -1.4  &   17.5 & 0.01 &  16.23 \\
                    &           &$4\times879$, &              & $3\times65$, &$2\times790$,& $2\times714$,&  30          & 30           &       &       &        &      &  \\ 
                    &           &$2\times854$  &              &              & 765         &    689       &              &              &       &      &        &      &  \\ 
\hline
NGC\,5272 (M\,3)$^{**}$& 2 &$6\times415$   & $4\times350$ &$4\times42$      &   &   & $4\times130$,& $4\times150$,&  -6.8   &    1.3   &   10.4 &0.01  & 15.07 \\
                    &           &               &               &              &             &              &  12          & 12          &       &       &        &      &  \\ 
\hline
NGC\,5286           &     2     & $2\times797$,& $4\times310$ &  65, 66      & 728, 603     & 655, 559     &$5\times350$,   & $5\times360$,  &   -0.7   &    -8.6  & 2.5&   0.24 &  16.08 \\ 
                    &           & $2\times668$ &              &              &              &              &      30        & 30 &       &      &        &      &  \\ 
\hline
\end{tabular}
\end{table}

\newpage
\begin{table}
\footnotesize
\begin{tabular}{l c |c c c |c c |c c |c c c c c}
\hline
\hline
                    &           & \multicolumn{3}{|c|}{Primary Fields} & \multicolumn{2}{|c|}{Parallel Fields} & \multicolumn{2}{|c|}{GO-10775 Fields} &        &       &       &          &  \\
Cluster             & \# Orbits & Exp.\,Time  & Exp.\,Time & Exp.\,Time & Exp.\,Time & Exp.\,Time & Exp.\,Time & Exp.\,Time &  X     & Y     & Z     & $E(B-V)$ & $(m-M)_V$ \\
                    &           & F275W[s]    & F336W[s]   & F438W[s]   &F475W[s]    & F814W[s]   & F606W[s]   & F814W[s]   & [kpc]  & [kpc] & [kpc] &          &           \\
\hline
NGC\,5466           &     4     &$2\times930$, & $4\times350$ & $2\times135$,&$2\times776$, & $2\times700$,& $5\times340$,& $5\times350$,& -5.0   &     3.0  &   15.7& 0.00 &   16.02 \\ 
                    &           &$2\times928$, &              & 138, 139     & 834, 835     & 763, 765     &  30          & 30           &       &       &        &      &  \\ 
                    &           &$4\times865$  &              &              &              &              &              &              &       &      &        &      &  \\ 
\hline
NGC\,5897           &     4     &$4\times926$, & $4\times350$ & $3\times140$,& 779, 781,    & $2\times761$,& $5\times340$,& $5\times350$,& 2.0   &    -3.2  &    6.7 &  0.09 &  15.76 \\
                    &           &$2\times875$, &              & 138          & 830, 833     & 710, 709     &  25          & 25           &       &       &        &      &  \\ 
                    &           &$2\times874$  &              &              &              &              &              &              &       &      &        &      &  \\ 
\hline
NGC\,5904 (M\,5)    &     2     & $2\times689$,& $4\times306$ &  58, 59      & 620, 621     & $2\times559$ &$4\times140$,   & $4\times140$,  & -3.2   &     0.3  &    5.9&  0.03 &  14.46 \\ 
                    &           & $2\times690$ &              &              &              &              &      7        & 7 &       &      &        &      &  \\ 
\hline
NGC\,5927           &     2     & $4\times668$ & $4\times310$ &  $2\times66$ & $2\times603$ & $2\times559$ &$5\times350$,   & $5\times350$,  & -1.9   &    -4.2  &    1.1&  0.45 &  15.82 \\ 
                    &           &              &              &              &              &              &      30        & 25 &       &      &        &      &  \\ 
\hline
NGC\,5986           &     2     &$2\times745$, & $4\times300$ & 60, 64     & 676, 603    & 603, 559 & $5\times350$,& $5\times350$,& 	1.0   &    -4.0  &    2.8 &  0.28 &  15.96 \\ 
                    &           &$2\times668$  &              &            &             &          &  20          & 20           &       &       &        &      &  \\ 
\hline
NGC\,6093 (M\,80)$^{**}$& 5 &$10\times885$   & $6\times657$ &$6\times85$  & $5\times760$,  & $5\times539$   & $5\times340$,& $5\times340$,&  1.1   &   -1.2   &    3.7& 0.18 &  15.56 \\
                    &           &               &               &              &   $5\times845$   &                &  10          & 10          &       &       &        &      &  \\ 
\hline
NGC\,6101           &     5     &$4\times851$, & $5\times415$ & $4\times165$,& $2\times762$,& $2\times686$,& $5\times370$,& $5\times380$,&2.7   &   -10.0  &   -3.8 &  0.05&   16.10 \\  
                    &           &$4\times889$, &              & 167          & $2\times800$,& $2\times724$,&  35          & 35           &       &       &        &      &  \\ 
                    &           &$2\times940$  &              &              &   851        & 775          &              &              &       &      &        &      &  \\ 
\hline
NGC\,6121 (M\,4)    &     2     &$2\times808$, & $4\times300$ & 66, 65       &  666, 739 & 593, 666 & $1\times1.5$, & $1\times1.5$,&-6.2   &    -0.3  &    1.0 &  0.35 & 12.82 \\
                    &           &$2\times735$  &              &              &           &          & $2\times25$,  & $4\times30$  &       &       &        &      &  \\ 
                    &           &              &              &              &           &          & $2\times30$   &              &       &      &        &      &  \\ 
NGC\,6121 (M\,4)$^*$& 2 &$10\times380$&            &              &           &         &   & &	& &   &   &  \\
\hline
NGC\,6144           &     2     &$4\times748$  & $4\times304$ & $2\times61$  & $2\times679$  & $2\times606$ & $5\times340$,& $5\times350$,& 0.2   &    -1.2  &    2.8 & 0.36  & 15.86 \\
                    &           &              &              &            &             &                  &  25          & 25           &       &       &        &      &  \\ 
\hline
NGC\,6171           &     4     &$4\times926$, & $4\times350$ & $2\times136$,& 800, 802, & $2\times761$,& $4\times130$,& $4\times150$,&	-2.4   &     0.3  &    2.9 &  0.33 &  15.05 \\
                    &           &$2\times896$, &              & 138, 140     & 830, 833  & 730, 731     &  12          & 12           &       &       &        &      &  \\ 
                    &           &$2\times895$  &              &              &           &              &              &              &       &      &        &      &  \\ 
\hline
NGC\,6205 (M\,13)$^{**}$& 2 &$6\times427$   & $4\times350$ &$4\times46$      &   &   & $4\times140$,& $4\times140$,& -5.5   &    4.6   &    5.1   & 0.02 &  14.33 \\
                    &           &               &               &              &             &              &  7          & 7          &       &       &        &      &  \\ 
\hline
NGC\,6218 (M\,12)   &     2     &$2\times790$,  & $4\times306$ & 58, 66     & 645, 721    & 572, 648     & $4\times90$,& $4\times90$,& -4.1   &     1.2  &    2.5 &  0.19 & 14.01 \\
                    &           &$2\times714$  &              &            &             &              &  4          & 4           &       &       &        &      &  \\ 
\hline
NGC\,6254 (M\,10)   &     2     &$2\times790$,  & $4\times306$ & 60, 66     & 644, 721    & 571, 648     & $4\times90$,& $4\times90$,& -4.4   &     1.1  &    2.1 & 0.28 &  14.08 \\
                    &           &$2\times713$  &              &            &             &              &  4          & 4           &       &       &        &      &  \\ 
\hline
NGC\,6304           &     2     &$2\times693$,  & $4\times304$ & 60, 61     & 624, 731    & 559, 658     & $5\times340$,& $5\times350$,& -2.5   &    -0.4  &    1.0 &  0.54 &  15.52 \\ 
                    &           &$2\times800$  &              &            &             &              &  20          & 20           &       &       &        &      &  \\ 
\hline
NGC\,6341  (M\,92)  &     2     &$2\times707$,  & $4\times304$ & 57, 59     & 638, 750    & 565, 677    & $4\times140$,& $4\times150$,& -5.8   &     6.3  &    5.1 & 0.02 &  14.65 \\ 
                    &           &$2\times819$  &              &            &             &              &  7          & 7           &       &       &        &      &  \\ 
\hline
\end{tabular}
\end{table}

\newpage
\begin{table}
\footnotesize
\begin{tabular}{l c |c c c |c c |c c |c c c c c}
\hline
\hline
                    &           & \multicolumn{3}{|c|}{Primary Fields} & \multicolumn{2}{|c|}{Parallel Fields} & \multicolumn{2}{|c|}{GO-10775 Fields} &        &       &       &          &  \\
Cluster             & \# Orbits & Exp.\,Time  & Exp.\,Time & Exp.\,Time & Exp.\,Time & Exp.\,Time & Exp.\,Time & Exp.\,Time &  X     & Y     & Z     & $E(B-V)$ & $(m-M)_V$ \\
                    &           & F275W[s]    & F336W[s]   & F438W[s]   &F475W[s]    & F814W[s]   & F606W[s]   & F814W[s]   & [kpc]  & [kpc] & [kpc] &          &           \\
\hline
NGC\,6352           &     2     &$2\times706$,  & $4\times311$ & 58, 72     & 637, 731    & 564, 658    & $4\times140$,& $4\times150$,& -3.0   &    -1.8  &   -0.3 & 0.22 &  14.43 \\
                    &           &$2\times800$  &              &            &             &              &  7          & 7           &       &       &        &      &  \\ 
\hline
NGC\,6362           &     2     &$2\times720$,  & $4\times323$ & 68, 67     & 651, 760    & 578, 687    & $4\times130$,& $4\times150$,& -2.3   &    -4.1  &   -1.9 &  0.09  & 14.68 \\
                    &           &$2\times829$  &              &            &             &              &  10          & 10           &       &       &        &      &  \\ 
\hline
NGC\,6366           &     2     &$2\times795$,  & $4\times305$ & $2\times62$& 644, 726    & 571, 653    & $4\times140$,& $4\times140$,& -5.1   &     1.1  &    1.4 &  0.71 &  14.94 \\ 
                    &           &$2\times713$  &              &            &             &              &  10          & 10           &       &       &        &      &  \\ 
\hline
NGC\,6388           &     4     & $2\times889$,& $4\times350$ & $2\times135$,& 793, 795, & 723, 724,   & $5\times340$,& $5\times350$,& 	1.3   &    -2.5  &   -0.8 & 0.37 &  16.13 \\
                    &           & $2\times888$,&              & $2\times133$ & 865, 906  & 796, 834    &      40      &     40       &       &      &        &      &  \\ 
                    &           & $2\times999$,&              &              &           &          &              &              &       &      &        &      &  \\ 
                    &           & $2\times961$ &              &              &              &              &              &              &       &      &        &      &  \\ 
\hline
NGC\,6397           &     2     &$2\times709$,  & $4\times310$ & $2\times66$& 640, 683    & 567, 610    & $4\times15$,& $4\times15$,& 	-6.2   &    -0.8  &   -0.1 & 0.18 &  12.37 \\
                    &           &$2\times752$  &              &            &             &              &  1          & 1           &       &       &        &      &  \\ 
\hline
NGC\,6441           &     4     & $2\times887$,& $4\times350$ & 123, 126,    & $2\times794$,& 722, 725,   & $5\times340$,& $5\times350$,& 3.2   &    -1.3  &   -0.6 & 0.47&   16.78 \\
                    &           & $2\times890$,&              & 128, 129    & 833, 835     & 763, 764    &      45      &     45       &       &      &        &      &  \\ 
                    &           & $2\times928$,&              &              &           &          &              &              &       &      &        &      &  \\ 
                    &           & $2\times929$ &              &              &              &              &              &              &       &      &        &      &  \\ 
\hline
NGC\,6496           &     2     &$2\times707$,  & $4\times303$ & 61, 73     & 638, 731    & 565, 658    & $10\times340$,& $5\times350$,& 2.6   &    -2.3  &   -1.6  &  0.15&   15.74\\ 
                    &           &$2\times800$  &              &            &             &              &  $2\times30$          & 30           &       &       &        &      &  \\ 
\hline
NGC\,6535           &     2     &$2\times713$,  & $4\times305$ & $2\times62$& 644, 724    & 571, 651    & $4\times130$,& $4\times150$,& -2.3   &     3.1  &    1.6 &   0.34  & 15.22 \\
                    &           &$2\times793$  &              &            &             &              &  12          & 12           &       &       &        &      &  \\ 
\hline
NGC\,6541           &     2     &$2\times708$,  & $4\times300$ & 65, 66     & 639, 689    & 566, 616    & $4\times140$,& $4\times150$,& -1.0   &    -1.4  &   -1.1 & 0.14 &  14.82 \\
                    &           &$2\times758$  &              &            &             &              &  8          & 8           &       &       &        &      &  \\ 
\hline
NGC\,6584           &     2     &$2\times709$,  & $4\times312$ & 62, 65     & 640, 726    & 567, 653    & $5\times350$,& $5\times360$,& 4.0   &    -4.0  &   -3.4 &  0.1 &   15.96 \\ 
                    &           &$2\times795$  &              &            &             &              &  25          & 25           &       &       &        &      &  \\ 
\hline
NGC\,6624           &     2     &$2\times707$,  & $4\times295$ & 62, 81     & 638, 731    & 565, 658    & $5\times350$,& $5\times350$,& -0.5   &     0.4  &   -0.7 & 0.28  & 15.36 \\
                    &           &$2\times800$  &              &            &             &              &  15          & 15           &       &       &        &      &  \\ 
\hline
NGC\,6637 (M\,69)   &     4     & $2\times887$,& $4\times350$ & 120, 122,    & 792, 794 & 722, 723,   & $5\times340$,& $5\times340$,& 0.3   &     0.3  &   -1.2&  0.18&   15.28\\ 
                    &           & $2\times888$,&              & 130, 131    & 827, 840     & 758, 768    &      18      &     18       &       &      &        &      &  \\ 
                    &           & $2\times923$,&              &              &           &          &              &              &       &      &        &      &  \\ 
                    &           & $2\times933$ &              &              &              &              &              &              &       &      &        &      &  \\ 
\hline
NGC\,6652           &     3     & $2\times690$,& $2\times305$,& 60, 69,      & 621, 707,  & 548, 633,    & $5\times340$,& $5\times340$,& 1.5   &     0.3  &   -1.6 & 0.09 &  15.28 \\
                    &           & $2\times775$,& $3\times313$,& 86           & 733        &  658         &      18      &     18       &       &      &        &      &  \\ 
                    &           & $2\times800$ &    312       &              &            &              &              &              &       &      &        &      &  \\ 
\hline
\end{tabular}
\end{table}

\newpage

\begin{table}
\footnotesize
\begin{tabular}{l c |c c c |c c |c c |c c c c c}
\hline
\hline
                    &           & \multicolumn{3}{|c|}{Primary Fields} & \multicolumn{2}{|c|}{Parallel Fields} & \multicolumn{2}{|c|}{GO-10775 Fields} &        &       &       &          &  \\
Cluster             & \# Orbits & Exp.\,Time  & Exp.\,Time & Exp.\,Time & Exp.\,Time & Exp.\,Time & Exp.\,Time & Exp.\,Time &  X     & Y     & Z     & $E(B-V)$ & $(m-M)_V$ \\
                    &           & F275W[s]    & F336W[s]   & F438W[s]   &F475W[s]    & F814W[s]   & F606W[s]   & F814W[s]   & [kpc]  & [kpc] & [kpc] &          &           \\
\hline
NGC\,6656 (M\,22)   &     1     &              & $3\times475$, & $2\times141$ &      &         & $4\times55$,& $5\times65$,& -5.1   &     0.6    &   -0.0 & 0.34 &  13.60 \\ 
                    &           &              &    476        &            &             &              &  3          & 3           &       &       &        &      &  \\ 
NGC\,6656 (M\,22)$^*$& 4 &$12\times800$&            &              & $4\times644$  & $4\times339$         &   & &	& &   &   &  \\
\hline
NGC\,6681 (M\,70)   &     2     &$2\times706$,  & $4\times294$ & 66, 83     & 637, 730    & 564, 658    & $4\times140$,& $4\times150$,&   0.5   &     0.4  &   -1.6 & 0.07 &  14.99 \\ 
                    &           &$2\times800$  &              &            &             &              &  10          & 10           &       &       &        &      &  \\ 
\hline
NGC\,6715 (M\,54)   &     6     & $3\times1755$,& $3\times433$,& $3\times170$,& $3\times370$,& $2\times1638$,& $4\times340$,& $4\times350$,&  17.3   &     2.5  &   -6.1 & 0.15  & 17.58 \\
                    &           & $2\times1920$,& $3\times475$ & $3\times190$ & $3\times390$ & 1469, 1473,   &      30      &     30       &       &      &        &      &  \\ 
                    &           & 1916          &              &              &              & 1474, 1634    &              &              &       &      &        &      &  \\ 
\hline
NGC\,6717           &     3     &$4\times686$,  & $5\times313$, & $3\times60$ & $2\times619$, & $2\times544$, & $4\times130$,& $4\times150$,&  -1.5   &     1.6  &   -0.9 & 0.22  & 14.94 \\
                    &           &$2\times687$  &    311          &            &    617         &   535           &  10          & 10           &       &       &        &      &  \\ 
\hline
NGC\,6723           &     3     &$4\times693$,  & $6\times313$ & $2\times61$, & 624, 626,  & $2\times551$, & $4\times140$,& $4\times150$,&  -0.0   &     0.0  &   -2.2 & 0.05   &14.84 \\
                    &           &$2\times734$  &              &       60     &    666         &   592           &  10          & 10           &       &       &        &      &  \\ 
\hline
NGC\,6752$^*$& 2 &$12\times360$&            &              & $2\times265$  &           & $4\times35$,& $4\times40$,& -5.0   &    -1.4    &   -1.3& 0.04 &  13.13\\ 
                    &           &  &              &           &             &              &  2         & 2           &       &       &        &      &  \\ 
\hline
NGC\,6779 (M\,56)   &     2     &$2\times706$,  & $4\times295$ & 64, 81      & 637, 731  & 564, 658 & $5\times340$,& $5\times350$,& -4.0   &     8.3  &    1.8 &  0.26 &  15.68 \\
                    &           &$2\times800$  &              &              &             &              &  20          & 20           &       &       &        &      &  \\ 
\hline
NGC\,6791           &     2     &$2\times700$,  & $2\times297$, & 65, 72      & 631, 638  & 559, 565 & & & -3.5   &    -1.7  &    1.4 &  0.12 &  13.43 \\ 
                    &           &$2\times707$  &  $2\times300$  &              &             &              &            &           &       &       &        &      &  \\ 
\hline
NGC\,6809 (M\,55)   &     2     &$2\times746$,  & $4\times294$ & 64, 66 & 677, 753  & 604, 680 & $4\times70$,& $4\times80$,& -3.4   &     0.8  &   -1.7 & 0.08 &  13.89 \\
                    &           &$2\times822$  &              &              &             &              &  4          & 4           &       &       &        &      &  \\ 
\hline
NGC\,6838 (M\,71)   &     2     &$2\times750$,  & $4\times303$ & $2\times65$ & 681, 723  & 608, 650 & $4\times75$,& $4\times80$,& -6.1   &     3.4  &    0.1 &0.25  & 13.80  \\
                    &           &$2\times792$  &              &              &             &              &  4          & 4           &       &       &        &      &  \\ 
\hline
NGC\,6934           &     2     &$2\times713$,  & $4\times304$ & 64, 70      & 644, 723  & 571, 650 & $5\times340$,& $5\times340$,& 0.8   &    11.7  &   -4.7 & 0.1 &   16.28 \\
                    &           &$2\times792$  &              &              &             &              &  45          & 45           &       &       &        &      &  \\ 
\hline
NGC\,6981 (M\,72)   &     2     &$2\times693$,  & $2\times304$, & 64, 70      & 624, 641  & 551, 568 & $4\times130$,& $4\times150$,& 	 3.4   &     8.3  &   -8.8 & 0.05 &  16.31 \\
                    &           &$2\times710$  &  $2\times305$  &              &             &              &  10          & 10          &       &       &        &      &  \\ 
\hline
NGC\,7078 (M\,15)$^{**}$& 3 &$3\times615$,   & $6\times350$ &$6\times65$      &   &   & $4\times130$,& $4\times150$,&  -4.4   &    8.3   &   -4.4 &0.1 &   15.39\\ 
                    &           &  $3\times700$           &               &              &             &              &  15          & 15         &       &       &        &      &  \\ 
\hline
NGC\,7089 (M\,2)    &     3     &$2\times676$,  & $6\times313$ & $2\times62$, & 611, 668,  & 534, 593, & $5\times340$,& $5\times340$,& 	-2.7   &     7.5  &   -6.3 &0.06  & 15.50  \\
                    &           &$2\times735$  &                &     70       &   717     &   643           &  20          & 20          &       &       &        &      &  \\ 
                    &           &$2\times785$  &          &              &             &              &            &          &       &       &        &      &  \\ 
\hline
NGC\,7099 (M\,30)   &     2     &$4\times725$   & $4\times303$ &$2\times65$      & $2\times656$  &  $2\times583$ & $4\times140$,& $4\times140$,& -3.4   &     2.5  &   -5.5 & 0.03 &  14.64 \\
                    &           &               &               &              &             &              &  7          & 7          &       &       &        &      &  \\ 
\hline
\end{tabular}

$^*{\phantom{1} }$ Data from GO--12311 \\
$^{**}$             Data from GO--12605

\end{table}

\end{landscape}


\begin{thebibliography}{}

\bibitem[Anderson et al.(2008)]{2008AJ....135.2055A} Anderson, J.,
  Sarajedini, A., Bedin, L.~R., King, I.~R., Piotto, G., Reid, I.~N.,
  Siegel, M., Majewski, S.~R., Paust, N.~E.~Q., Aparicio, A., Milone,
  A.~P., Chaboyer, B., \& Rosenberg, A.\ 2008, \aj, 135, 2055

\bibitem[Anderson \& van der Marel(2010)]{2010ApJ...710.1032A}
  Anderson, J., \& van der Marel, R.~P.\ 2010, \apj, 710, 1032

\bibitem[Bastian, Cabrera-Ziri, Davies, \&
  Larsen(2013)]{2013MNRAS.436.2852B} Bastian, N., Cabrera-Ziri, I.,
  Davies, B., \& Larsen, S.~S.\ 2013a, \mnras, 436, 2852

\bibitem[Bastian et al.(2013)]{2013MNRAS.436.2398B} Bastian, N.,
  Lamers, H.~J.~G.~L.~M., de Mink, S.~E., Longmore, S.~N., Goodwin,
  S.~P., \& Gieles, M.\ 2013b, \mnras, 436, 2398

\bibitem[Barbuy et al.(1998)]{1998A&A...333..117B} Barbuy, B., Bica,
  E., \& Ortolani, S.\ 1998, \aap, 333, 117

\bibitem[Bedin et al.(2004)]{2004ApJ...605L.125B} Bedin, L.~R.,
  Piotto, G., Anderson, J., Cassisi, S., King, I.~R., Momany, Y., \&
  Carraro, G.\ 2004, \apjl, 605, L125

\bibitem[Bedin et al.(2008)]{2008ApJ...678.1279B} Bedin, L.~R., King,
  I.~R., Anderson, J., Piotto, G., Salaris, M., Cassisi, S., \&
  Serenelli, A.\ 2008, \apj, 678, 1279

\bibitem[Bekki(2011)]{2011MNRAS.412.2241B} Bekki, K.\ 2011, \mnras,
  412, 2241

\bibitem[Bellini et al.(2009)]{2009A&A...507.1393B} Bellini, A.,
  Piotto, G., Bedin, L.~R., King, I.~R., Anderson, J., Milone, A.~P.,
  \& Momany, Y.\ 2009, \aap, 507, 1393

\bibitem[Bellini et al.(2010)]{2010AJ....140..631B} Bellini, A.,
  Bedin, L.~R., Piotto, G., Milone, A.~P., Marino, A.~F., \&
  Villanova, S.\ 2010, \aj, 140, 631
 
\bibitem[Bellini et al.(2013a)]{2013ApJ...765...32B} Bellini, A.,
  Piotto, G., Milone, A.~P., King, I.~R., Renzini, A., Cassisi, S.,
  Anderson, J., Bedin, L.~R., Nardiello, D., Pietrinferni, A., \&
  Sarajedini, A.\ 2013a, \apj, 765, 32

\bibitem[Bellini et al.(2013b)]{2013MmSAI..84..140B} Bellini, A., van
  der Marel, R.~P., \& Anderson, J.\ 2013b, \memsai, 84, 140

\bibitem[Bellini et al.(2013c)]{2013ApJ...769L..32B} Bellini, A.,
  Anderson, J., Salaris, M., et al.\ 2013c, \apjl, 769, L32

\bibitem[Bellini et al.(2014)]{2014Bellsubmitted} Bellini A. et al.\
  2014, submitted to ApJ

\bibitem[Brown et al.(2001)]{2001ApJ...562..368B} Brown, T.~M.,
  Sweigart, A.~V., Lanz, T., Landsman, W.~B., \& Hubeny, I.\ 2001,
  \apj, 562, 368

\bibitem[Brown et al.(2010)]{2010ApJ...718.1332B} Brown, T.~M.,
  Sweigart, A.~V., Lanz, T., Smith, E., Landsman, W.~B., \& Hubeny,
  I.\ 2010, \apj, 718, 1332

\bibitem[Busso et al.(2007)]{2007A&A...474..105B} Busso, G., Cassisi,
  S., Piotto, G., Castellani, M., Romaniello, M., Catelan, M.,
  Djorgovski, S.~G., Recio Blanco, A., Renzini, A., Rich, M.~R.,
  Sweigart, A.~V., \& Zoccali, M.\ 2007, \aap, 474, 105

\bibitem[Caloi \& D'Antona(2007)]{2007A&A...463..949C} Caloi, V., \&
  D'Antona, F.\ 2007, \aap, 463, 949

\bibitem[Campbell et al.(2010)]{2010MmSAI..81.1004C} Campbell, S.~W.,
  Yong, D., Wylie-de Boer, E.~C., Stancliffe, R.~J., Lattanzio, J.~C.,
  Angelou, G.~C., Grundahl, F., \& Sneden, C.\ 2010, \memsai, 81, 1004

\bibitem[Campbell et al.(2013)]{2013Natur.498..198C} Campbell, S.~W.,
  D'Orazi, V., Yong, D., et al.\ 2013, \nat, 498, 198

\bibitem[Carretta et al.(2009)]{2009A&A...505..117C} Carretta, E.,
  Bragaglia, A., Gratton, R.~G., Lucatello, S., Catanzaro, G., Leone,
  F., Bellazzini, M., Claudi, R., D'Orazi, V., Momany, Y., Ortolani,
  S., Pancino, E., Piotto, G., Recio-Blanco, A., \& Sabbi, E.\ 2009,
  \aap, 505, 117

\bibitem[Carretta et al.(2010)]{2010A&A...520A..95C} Carretta, E.,
  Bragaglia, A., Gratton, R.~G., Lucatello, S., Bellazzini, M.,
  Catanzaro, G., Leone, F., Momany, Y., Piotto, G., \& D'Orazi, V.\
  2010, \aap, 520, A95

\bibitem[Cassisi et al.(2008)]{2008ApJ...672L.115C} Cassisi, S.,
  Salaris, M., Pietrinferni, A., Piotto, G., Milone, A.~P., Bedin,
  L.~R., \& Anderson, J.\ 2008, \apjl, 672, L115

\bibitem[Cassisi et al.(2011)]{2011A&A...527A..59C} Cassisi, S.,
  Mar{\'{\i}}n-Franch, A., Salaris, M., Aparicio, A., Monelli, M., \&
  Pietrinferni, A.\ 2011, \aap, 527, A59

\bibitem[Cassisi et al.(2013)]{2013A&A...554A..19C} Cassisi, S.,
  Mucciarelli, A., Pietrinferni, A., Salaris, M., \& Ferguson, J.\
  2013, \aap, 554, A19

\bibitem[Cassisi \& Salaris(2014)]{2014A&A...563A..10C} Cassisi, S.,
  \& Salaris, M.\ 2014, \aap, 563, A10

\bibitem[Cote(1999)]{1999AJ....118..406C} Cot\'e, P.\ 1999, \aj, 118,
  406

\bibitem[Da Costa, Held, \& Saviane(2014)]{2014MNRAS.438.3507D} Da
  Costa, G.~S., Held, E.~V., \& Saviane, I.\ 2014, \mnras, 438, 3507


\bibitem[Dalessandro et al.(2011)]{2011MNRAS.410..694D} Dalessandro,
  E., Salaris, M., Ferraro, F.~R., Cassisi, S., Lanzoni, B., Rood,
  R.~T., Fusi Pecci, F., \& Sabbi, E.\ 2011, \mnras, 410, 694

\bibitem[Dalessandro et al.(2013)]{2013MNRAS.430..459D} Dalessandro,
  E., Salaris, M., Ferraro, F.~R., Mucciarelli, A., \& Cassisi, S.\
  2013a, \mnras, 430, 459

\bibitem[Dalessandro et al.(2013)]{2013ApJ...778..135D} Dalessandro,
  E., Ferraro, F.~R., Massari, D., et al.\ 2013b, \apj, 778, 135

\bibitem[Dantona, Gratton, \& Chieffi(1983)]{1983MmSAI..54..173D}
  Dantona, F., Gratton, R., \& Chieffi, A.\ 1983, \memsai, 54, 173

\bibitem[D'Antona et al.(2002)]{2002A&A...395...69D} D'Antona, F.,
  Caloi, V., Montalb{\'a}n, J., Ventura, P., \& Gratton, R.\ 2002,
  \aap, 395, 69

\bibitem[D'Antona \& Caloi(2004)]{2004ApJ...611..871D} D'Antona, F.,
  \& Caloi, V.\ 2004, \apj, 611, 871

\bibitem[D'Antona \& Caloi(2008)]{2008MNRAS.390..693D} D'Antona, F.,
  \& Caloi, V.\ 2008, \mnras, 390, 693

\bibitem[D'Antona et al.(2014)]{2014MNRAS.443.3302D} D'Antona, F.,
  Ventura, P., Decressin, T., Vesperini, E., \& D'Ercole, A.\ 2014,
  \mnras, 443, 3302

\bibitem[D'Cruz, Dorman, Rood, \&
  O'Connell(1996)]{1996ApJ...466..359D} D'Cruz, N.~L., Dorman, B.,
  Rood, R.~T., \& O'Connell, R.~W.\ 1996, \apj, 466, 359

\bibitem[Decressin et al.(2007)]{2007A&A...464.1029D} Decressin, T.,
  Meynet, G., Charbonnel, C., Prantzos, N., \& Ekstr{\"o}m, S.\ 2007,
  \aap, 464, 1029

\bibitem[Decressin, Baumgardt, Charbonnel, \&
  Kroupa(2010)]{2010A&A...516A..73D} Decressin, T., Baumgardt, H.,
  Charbonnel, C., \& Kroupa, P.\ 2010, \aap, 516, A73

\bibitem[de Mink, Pols, Langer, \& Izzard(2009)]{2009A&A...507L...1D}
  de Mink, S.~E., Pols, O.~R., Langer, N., \& Izzard, R.~G.\ 2009,
  \aap, 507, L1

\bibitem[D'Ercole et al.(2008)]{2008MNRAS.391..825D} D'Ercole, A.,
  Vesperini, E., D'Antona, F., McMillan, S.~L.~W., \& Recchi, S.\
  2008, \mnras, 391, 825

\bibitem[D'Ercole, D'Antona, \& Vesperini(2011)]{2011MNRAS.415.1304D}
  D'Ercole, A., D'Antona, F., \& Vesperini, E.\ 2011, \mnras, 415,
  1304

\bibitem[Djorgovski, Piotto, \& Capaccioli(1993)]{1993AJ....105.2148D}
  Djorgovski, S., Piotto, G., \& Capaccioli, M.\ 1993, \aj, 105, 2148

\bibitem[Ferraro, Possenti, D'Amico, \&
  Sabbi(2001)]{2001ApJ...561L..93F} Ferraro, F.~R., Possenti, A.,
  D'Amico, N., \& Sabbi, E.\ 2001, \apjl, 561, L93

\bibitem[Ferraro et al.(2003)]{2003ASPC..296..143F} Ferraro, F.~R.,
  Possenti, A., Lagani, P., Sabbi, E., D'Amico, N., \& Rood, R.~T.\
  2003a, New Horizons in Globular Cluster Astronomy, 296, 143

\bibitem[Ferraro, Possenti, Sabbi, \&
  D'Amico(2003)]{2003ApJ...596L.211F} Ferraro, F.~R., Possenti, A.,
  Sabbi, E., \& D'Amico, N.\ 2003b, \apjl, 596, L211

\bibitem[Ferraro et al.(2009)]{2009Natur.462..483F} Ferraro, F.~R.,
  Dalessandro, E., Mucciarelli, A., Beccari, G., Rich, R.~M., Origlia,
  L., Lanzoni, B., Rood, R.~T., Valenti, E., Bellazzini, M., Ransom,
  S.~M., \& Cocozza, G.\ 2009a, \nat, 462, 483

\bibitem[Ferraro et al.(2009)]{2009Natur.462.1028F} Ferraro, F.~R.,
  Beccari, G., Dalessandro, E., Lanzoni, B., Sills, A., Rood, R.~T.,
  Pecci, F.~F., Karakas, A.~I., Miocchi, P., \& Bovinelli, S.\ 2009b,
  \nat, 462, 1028

\bibitem[Ferraro et al.(2012)]{2012Natur.492..393F} Ferraro, F.~R.,
  Lanzoni, B., Dalessandro, E., Beccari, G., Pasquato, M., Miocchi,
  P., Rood, R.~T., Sigurdsson, S., Sills, A., Vesperini, E., Mapelli,
  M., Contreras, R., Sanna, N., \& Mucciarelli, A.\ 2012, \nat, 492,
  393

\bibitem[Geisler et al.(2012)]{2012ApJ...756L..40G} Geisler, D.,
  Villanova, S., Carraro, G., Pilachowski, C., Cummings, J., Johnson,
  C.~I., \& Bresolin, F.\ 2012, \apjl, 756, L40

\bibitem[Gratton et al.(2011)]{2011A&A...534A.123G} Gratton, R.~G.,
  Lucatello, S., Carretta, E., Bragaglia, A., D'Orazi, V., \& Momany,
  Y.~A.\ 2011, \aap, 534, A123

\bibitem[Gratton et al.(2012)]{2012A&A...539A..19G} Gratton, R.~G.,
  Lucatello, S., Carretta, E., Bragaglia, A., D'Orazi, V., Al Momany,
  Y., Sollima, A., Salaris, M., \& Cassisi, S.\ 2012, \aap, 539, A19

\bibitem[Gratton et al.(2013)]{2013A&A...549A..41G} Gratton, R.~G.,
  Lucatello, S., Sollima, A., Carretta, E., Bragaglia, A., Momany, Y.,
  D'Orazi, V., Cassisi, S., Pietrinferni, A., \& Salaris, M.\ 2013,
  \aap, 549, A41

\bibitem[Harris(1996)]{1996AJ....112.1487H} Harris, W.~E.\ 1996, \aj,
  112, 1487

\bibitem[Kaviraj et al.(2007)]{2007ApJS..173..619K} Kaviraj, S.,
  Schawinski, K., Devriendt, J.~E.~G., Ferreras, I., Khochfar, S.,
  Yoon, S.-J., Yi, S.~K., Deharveng, J.-M., Boselli, A., Barlow, T.,
  Conrow, T., Forster, K., Friedman, P.~G., Martin, D.~C., Morrissey,
  P., Neff, S., Schiminovich, D., Seibert, M., Small, T., Wyder, T.,
  Bianchi, L., Donas, J., Heckman, T., Lee, Y.-W., Madore, B.,
  Milliard, B., Rich, R.~M., \& Szalay, A.\ 2007, \apjs, 173, 619

\bibitem[King et al.(2005)]{2005AJ....130..626K} King, I.~R., Bedin,
  L.~R., Piotto, G., Cassisi, S., \& Anderson, J.\ 2005, \aj, 130, 626

\bibitem[King et al.(2012)]{2012AJ....144....5K} King, I.~R., Bedin,
  L.~R., Cassisi, S., Milone, A.~P., Bellini, A., Piotto, G.,
  Anderson, J., Pietrinferni, A., \& Cordier, D.\ 2012, \aj, 144, 5

\bibitem[Lanzoni et al.(2013)]{2013ApJ...769..107L} Lanzoni, B.,
  Mucciarelli, A., Origlia, L., et al.\ 2013, \apj, 769, 107

\bibitem[Lardo et al.(2011)]{2011A&A...525A.114L} Lardo, C.,
  Bellazzini, M., Pancino, E., Carretta, E., Bragaglia, A., \&
  Dalessandro, E.\ 2011, \aap, 525, A114

\bibitem[L{\"u}tzgendorf et al.(2013)]{2013A&A...552A..49L}
  L{\"u}tzgendorf, N., Kissler-Patig, M., Gebhardt, K., et al.\ 2013,
  \aap, 552, A49

\bibitem[MacKenty \& WFC3 Team(2012)]{2012AAS...22013609M} MacKenty,
  J.~W., \& WFC3 Team 2012, American Astronomical Society Meeting
  Abstracts \#220, 220, \#136.09

\bibitem[Mar{\'{\i}}n-Franch et al.(2009)]{2009ApJ...694.1498M}
  Mar{\'{\i}}n-Franch, A., Aparicio, A., Piotto, G., Rosenberg, A.,
  Chaboyer, B., Sarajedini, A., Siegel, M., Anderson, J., Bedin,
  L.~R., Dotter, A., Hempel, M., King, I., Majewski, S., Milone,
  A.~P., Paust, N., \& Reid, I.~N.\ 2009, \apj, 694, 1498

\bibitem[Marino et al.(2008)]{2008A&A...490..625M} Marino, A.~F.,
  Villanova, S., Piotto, G., Milone, A.~P., Momany, Y., Bedin, L.~R.,
  \& Medling, A.~M.\ 2008, \aap, 490, 625

\bibitem[Marino et al.(2009)]{2009A&A...505.1099M} Marino, A.~F.,
  Milone, A.~P., Piotto, G., Villanova, S., Bedin, L.~R., Bellini, A.,
  \& Renzini, A.\ 2009, \aap, 505, 1099

\bibitem[Marino et al.(2011)]{2011ApJ...730L..16M} Marino, A.~F.,
  Villanova, S., Milone, A.~P., Piotto, G., Lind, K., Geisler, D., \&
  Stetson, P.~B.\ 2011a, \apjl, 730, L16

\bibitem[Marino et al.(2011)]{2011ApJ...731...64M} Marino, A.~F., Milone, 
A.~P., Piotto, G., Villanova, S., Gratton, R., D'Antona, F., Anderson, J., 
Bedin, L.~R., Bellini, A., Cassisi, S., Geisler, D., Renzini, A., 
\& Zoccali, M.\ 2011b, \apj, 731, 64 

\bibitem[Marino et al.(2011)]{2011A&A...532A...8M} Marino, A.~F.,
  Sneden, C., Kraft, R.~P., Wallerstein, G., Norris, J.~E., da Costa,
  G., Milone, A.~P., Ivans, I.~I., Gonzalez, G., Fulbright, J.~P.,
  Hilker, M., Piotto, G., Zoccali, M., \& Stetson, P.~B.\ 2011c, \aap,
  532, A8

\bibitem[Marino et al.(2012)]{2012ApJ...746...14M} Marino, A.~F.,
  Milone, A.~P., Piotto, G., Cassisi, S., D'Antona, F., Anderson, J.,
  Aparicio, A., Bedin, L.~R., Renzini, A., \& Villanova, S.\ 2012,
  \apj, 746, 14

\bibitem[Marino, Milone, \& Lind(2013)]{2013ApJ...768...27M} Marino,
  A.~F., Milone, A.~P., \& Lind, K.\ 2013, \apj, 768, 27

\bibitem[Marino et al.(2014)]{2014MNRAS.437.1609M} Marino, A.~F., Milone, 
A.~P., Przybilla, N., Bergemann, M., Lind, K., Asplund, M., Cassisi, S., 
Catelan, M., Casagrande, L., Valcarce, A.~A.~R., Bedin, L.~R., Cort{\'e}s, 
C., D'Antona, F., Jerjen, H., Piotto, G., Schlesinger, K., Zoccali, M., 
\& Angeloni, R.\ 2014, \mnras, 437, 1609 

\bibitem[Milone et al.(2010)]{2010ApJ...709.1183M} Milone, A.~P.,
  Piotto, G., King, I.~R., Bedin, L.~R., Anderson, J., Marino, A.~F.,
  Momany, Y., Malavolta, L., \& Villanova, S.\ 2010, \apj, 709, 1183

\bibitem[Milone et al.(2012)]{2012ApJ...744...58M} Milone, A.~P.,
  Piotto, G., Bedin, L.~R., King, I.~R., Anderson, J., Marino, A.~F.,
  Bellini, A., Gratton, R., Renzini, A., Stetson, P.~B., Cassisi, S.,
  Aparicio, A., Bragaglia, A., Carretta, E., D'Antona, F., Di
  Criscienzo, M., Lucatello, S., Monelli, M., \& Pietrinferni, A.\
  2012a, \apj, 744, 58

\bibitem[Milone et al.(2012)]{2012ApJ...745...27M} Milone, A.~P.,
  Marino, A.~F., Piotto, G., Bedin, L.~R., Anderson, J., Aparicio, A.,
  Cassisi, S., \& Rich, R.~M.\ 2012b, \apj, 745, 27

\bibitem[Milone et al.(2012)]{2012A&A...540A..16M} Milone, A.~P.,
  Piotto, G., Bedin, L.~R., Aparicio, A., Anderson, J., Sarajedini,
  A., Marino, A.~F., Moretti, A., Davies, M.~B., Chaboyer, B., Dotter,
  A., Hempel, M., Mar{\'{\i}}n-Franch, A., Majewski, S., Paust,
  N.~E.~Q., Reid, I.~N., Rosenberg, A., \& Siegel, M.\ 2012c, \aap,
  540, A16

\bibitem[Milone et al.(2012)]{2012A&A...537A..77M} Milone, A.~P.,
  Piotto, G., Bedin, L.~R., Cassisi, S., Anderson, J., Marino, A.~F.,
  Pietrinferni, A., \& Aparicio, A.\ 2012d, \aap, 537, A77

\bibitem[Milone et al.(2013)]{2013ApJ...767..120M} Milone, A.~P.,
  Marino, A.~F., Piotto, G., Bedin, L.~R., Anderson, J., Aparicio, A.,
  Bellini, A., Cassisi, S., D'Antona, F., Grundahl, F., Monelli, M.,
  \& Yong, D.\ 2013, \apj, 767, 120

\bibitem[Milone et al.(2014)]{2014MNRAS.439.1588M} Milone, A.~P.,
  Marino, A.~F., Bedin, L.~R., Piotto, G., Cassisi, S., Dieball, A.,
  Anderson, J., Jerjen, H., Asplund, M., Bellini, A., Brogaard, K.,
  Dotter, A., Giersz, M., Heggie, D.~C., Knigge, C., Rich, R.~M., van
  den Berg, M., \& Buonanno, R.\ 2014a, \mnras, 439, 1588

\bibitem[Milone et al.(2014)]{2014ApJ...785...21M} Milone, A.~P.,
  Marino, A.~F., Dotter, A., Norris, J.~E., Jerjen, H., Piotto, G.,
  Cassisi, S., Bedin, L.~R., Recio Blanco, A., Sarajedini, A.,
  Asplund, M., Monelli, M., \& Aparicio, A.\ 2014b, \apj, 785, 21

\bibitem[Miocchi et al.(2013)]{2013ApJ...774..151M} Miocchi, P.,
  Lanzoni, B., Ferraro, F.~R., Dalessandro, E., Vesperini, E.,
  Pasquato, M., Beccari, G., Pallanca, C., \& Sanna, N.\ 2013, \apj,
  774, 151

\bibitem[Norris, Cottrell, Freeman, \& Da
  Costa(1981)]{1981ApJ...244..205N} Norris, J., Cottrell, P.~L.,
  Freeman, K.~C., \& Da Costa, G.~S.\ 1981, \apj, 244, 205

\bibitem[Norris \& Da Costa(1995)]{1995ApJ...447..680N} Norris, J.~E.,
  \& Da Costa, G.~S.\ 1995, \apj, 447, 680

\bibitem[Norris(2004)]{2004ApJ...612L..25N} Norris, J.~E.\ 2004,
  \apjl, 612, L25

\bibitem[Noyola et al.(2010)]{2010ApJ...719L..60N} Noyola, E.,
  Gebhardt, K., Kissler-Patig, M., L\"utzgendorf, N., Jalali, B., de
  Zeeuw, P. T., Baumgardt, H. 2010, \apj, 719, L60

\bibitem[Pallanca et al.(2010)]{2010ApJ...725.1165P} Pallanca, C.,
  Dalessandro, E., Ferraro, F.~R., Lanzoni, B., Rood, R.~T., Possenti,
  A., D'Amico, N., Freire, P.~C., Stairs, I., Ransom, S.~M., \&
  B{\'e}gin, S.\ 2010, \apj, 725, 1165

\bibitem[Paust et al.(2010)]{2010AJ....139..476P} Paust, N.~E.~Q.,
  Reid, I.~N., Piotto, G., Aparicio, A., Anderson, J., Sarajedini, A.,
  Bedin, L.~R., Chaboyer, B., Dotter, A., Hempel, M., Majewski, S.,
  Mar{\'{\i}}n-Franch, A., Milone, A., Rosenberg, A., \& Siegel, M.\
  2010, \aj, 139, 476

\bibitem[Pilachowski, Sneden, Kraft, \&
  Langer(1996)]{1996AAS...188.0301P} Pilachowski, C., Sneden, C.,
  Kraft, R.~P., \& Langer, G.~E.\ 1996, Bulletin of the American
  Astronomical Society, 28, 821

\bibitem[Piotto et al.(2004)]{2004ApJ...604L.109P} Piotto, G., De
  Angeli, F., King, I.~R., Djorgovski, S.~G., Bono, G., Cassisi, S.,
  Meylan, G., Recio-Blanco, A., Rich, R.~M., \& Davies, M.~B.\ 2004,
  \apjl, 604, L109

\bibitem[Piotto et al.(2005)]{2005ApJ...621..777P} Piotto, G.,
  Villanova, S., Bedin, L.~R., Gratton, R., Cassisi, S., Momany, Y.,
  Recio-Blanco, A., Lucatello, S., Anderson, J., King, I.~R.,
  Pietrinferni, A., \& Carraro, G.\ 2005, \apj, 621, 777

\bibitem[Piotto et al.(2007)]{2007ApJ...661L..53P} Piotto, G., Bedin,
  L.~R., Anderson, J., King, I.~R., Cassisi, S., Milone, A.~P.,
  Villanova, S., Pietrinferni, A., \& Renzini, A.\ 2007, \apjl, 661,
  L53

\bibitem[Piotto et al.(2012)]{2012ApJ...760...39P} Piotto, G., Milone,
  A.~P., Anderson, J., Bedin, L.~R., Bellini, A., Cassisi, S., Marino,
  A.~F., Aparicio, A., \& Nascimbeni, V.\ 2012, \apj, 760, 39

\bibitem[Piotto et al.(2013)]{2013ApJ...775...15P} Piotto, G., Milone,
  A.~P., Marino, A.~F., Bedin, L.~R., Anderson, J., Jerjen, H.,
  Bellini, A., \& Cassisi, S.\ 2013, \apj, 775, 15

\bibitem[Renzini(2008)]{2008MNRAS.391..354R} Renzini, A.\ 2008,
  \mnras, 391, 354

\bibitem[Renzini(2013)]{2013MmSAI..84..162R} Renzini, A.\ 2013,
  \memsai, 84, 162

\bibitem[Rich et al.(1997)]{1997ApJ...484L..25R} Rich, R.~M., Sosin,
  C., Djorgovski, S.~G., Piotto, G., King, I.~R., Renzini, A.,
  Phinney, E.~S., Dorman, B., Liebert, J., \& Meylan, G.\ 1997, \apjl,
  484, L25

\bibitem[Richer et al. (2013)]{2013ApJ...771L..15R} Richer, H.~B.,
  Heyl, J., Anderson, J., Kalirai, J.~S., Shara, M.~M., Dotter, A.,
  Fahlman, G.~G., Rich, R.~M.\ 2013, \apjl, 771, L15

\bibitem[Romani \& Weinberg(1991)]{1991ASPC...13..443R} Romani, R.~W.,
  \& Weinberg, M.~D.\ 1991, The Formation and Evolution of Star
  Clusters, 13, 443

\bibitem[Rood\& Crocker(1989)]{1989upsf.conf..103R} Rood, R.~T., \&
  Crocker, D.~A.\ 1989, IAU Colloq.~111: The Use of pulsating stars in
  fundamental problems of astronomy, 103

\bibitem[Salaris \& Cassisi(2014)]{2014A&A...566A.109S} Salaris, M.,
  \& Cassisi, S.\ 2014, \aap, 566, A109

\bibitem[Schaerer \& Charbonnel(2011)]{2011MNRAS.413.2297S} Schaerer,
  D., \& Charbonnel, C.\ 2011, \mnras, 413, 2297

\bibitem[Sbordone, Salaris, Weiss, \&
  Cassisi(2011)]{2011A&A...534A...9S} Sbordone, L., Salaris, M.,
  Weiss, A., \& Cassisi, S.\ 2011, \aap, 534, A9

\bibitem[Schlegel, Finkbeiner,\& Davis(1998)]{1998ApJ...500..525S}
  Schlegel, D.~J., Finkbeiner, D.~P., \& Davis, M.\ 1998, \apj, 500,
  525

\bibitem[Sollima et al.(2007)]{2007ApJ...654..915S} Sollima, A.,
  Ferraro, F.~R., Bellazzini, M., Origlia, L., Straniero, O., \&
  Pancino, E.\ 2007, \apj, 654, 915

\bibitem[Trenti \& van der Marel(2013)]{2013MNRAS.435.3272T} Trenti,
  M., \& van der Marel, R.\ 2013, \mnras, 435, 3272

\bibitem[Vanbeveren, Mennekens,\& De Greve(2012)]{2012A&A...543A...4V}
  Vanbeveren, D., Mennekens, N., \& De Greve, J.~P.\ 2012, \aap, 543,
  A4

\bibitem[van der Marel \& Anderson(2010)]{2010ApJ...710.1063V} van der
  Marel, R.~P., \& Anderson, J.\ 2010, \apj, 710, 1063

\bibitem[Ventura et al.(2009)]{2009MNRAS.399..934V} Ventura, P.,
  Caloi, V., D'Antona, F., Ferguson, J., Milone, A., \& Piotto, G.~P.\
  2009, \mnras, 399, 934

\bibitem[Ventura et al.(2014)]{2014MNRAS.437.3274V} Ventura, P.,
  Criscienzo, M.~D., D'Antona, F., Vesperini, E., Tailo, M.,
  Dell'Agli, F., \& D'Ercole, A.\ 2014, \mnras, 437, 3274

\bibitem[Vesperini, McMillan, D'Antona, \&
  D'Ercole(2010)]{2010ApJ...718L.112V} Vesperini, E., McMillan,
  S.~L.~W., D'Antona, F., \& D'Ercole, A.\ 2010, \apjl, 718, L112

\bibitem[Vesperini, McMillan, D'Antona, \&
  D'Ercole(2013)]{2013MNRAS.429.1913V} Vesperini, E., McMillan,
  S.~L.~W., D'Antona, F., \& D'Ercole, A.\ 2013, \mnras, 429, 1913

\bibitem[Villanova, Geisler, Piotto, \&
  Gratton(2012)]{2012ApJ...748...62V} Villanova, S., Geisler, D.,
  Piotto, G., \& Gratton, R.~G.\ 2012, \apj, 748, 62

\bibitem[Vink{\'o} et al.(2009)]{2009ApJ...695..619V} Vink{\'o}, J.,
  S{\'a}rneczky, K., Balog, Z., Immler, S., Sugerman, B.~E.~K., Brown,
  P.~J., Misselt, K., Szab{\'o}, G.~M., Csizmadia, S., Kun, M.,
  Klagyivik, P., Foley, R.~J., Filippenko, A.~V., Cs{\'a}k, B., \&
  Kiss, L.~L.\ 2009, \apj, 695, 619

\bibitem[Yong \& Grundahl(2008)]{2008ApJ...672L..29Y} Yong, D., \&
  Grundahl, F.\ 2008, \apjl, 672, L29

\bibitem[Yong et al.(2008)]{2008ApJ...684.1159Y} Yong, D., Grundhal,
  F., Johnson, J.~A., \& Asplund, M.\ 2008, \apj, 684, 1159

\bibitem[Yong et al.(2014)]{2014MNRAS.441.3396Y} Yong, D., Roederer,
  I.~U., Grundahl, F., Da Costa, G.~S., Karakas, A.~I., Norris, J.~E.,
  Aoki, W., Fishlock, C.~K., Marino, A.~F., Milone, A.~P., \&
  Shingles, L.~J.\ 2014, \mnras, 441, 3396

\bibitem[Villanova et al.(2013)]{ 2013ApJ...778..186V} Villanova, S.,
  Geisler, D., Carraro, G., Moni Bidin, C., \& Muñoz, C.\ 2013, \apj,
  778, 186

\end{thebibliography}
\end{document}